\documentclass[aps,prd,eqsecnum,12pt,showpacs,preprintnumbers,nofootinbib]{revtex4-2}
\usepackage{geometry}                
\usepackage{graphicx}
\usepackage[font=small,format=plain,labelfont=bf,up,textfont=normal,up,justification=justified,singlelinecheck=false]{caption}
\usepackage{caption,subcaption}
\usepackage{float,placeins,relsize,nccmath}

\usepackage{float}
\usepackage{amssymb,amsmath,amsfonts,amsopn,bm,dsfont,mathrsfs}
\usepackage{mathtools}
\allowdisplaybreaks[4]

\usepackage{empheq}

\usepackage[mathscr]{euscript}
\usepackage{upgreek}
\usepackage{relsize}
\DeclareMathAlphabet{\mathpzc}{OT1}{pzc}{m}{it}

\newcommand{\mqs}{\mspace{-1.5mu}}

\newcommand{\T}{\scalebox{1.36}{$\uptau$}}
\usepackage[colorlinks,linkcolor=blue, pdfstartview=FitH]{hyperref}
\usepackage{enumitem}
\usepackage{txfonts}
\usepackage{wrapfig}
\usepackage[compact,small]{titlesec}

\usepackage{amsbsy}

\hypersetup{linkcolor=blue,citecolor=blue,filecolor=black,urlcolor=blue}

\IfFileExists{dsfont.sty}
	{\usepackage{dsfont}
         \let\mathbb=\mathds
         \newcommand{\id}{\mathds{1}}}
	{\typeout{Package dsfont.sty was not found, using alternative macros.}
         \let\mathds=\mathbb
         \newcommand{\id}{\mbox{1 \kern-.59em \textrm{l}}}}

\renewcommand\a{\alpha}
\renewcommand\b{\beta}
\renewcommand\d{\delta}
\newcommand\ka{\kappa}
\renewcommand\l{\lambda}
\renewcommand\r{\rho}
\renewcommand\i{\iota}
\renewcommand\t{\tau}

\renewcommand\th{\theta}

\newcommand\e{\epsilon}
\newcommand\g{\gamma}

\newcommand\m{\mu}
\newcommand\n{\nu}
\newcommand\x{\xi}
\newcommand\p{\pi}
\newcommand\h{\eta}
\newcommand\s{\sigma}
\newcommand\f{\phi}

\newcommand\vf{\raisebox{1pt}{$\varphi$}}

\newcommand\La{\Lambda}
\renewcommand\P{\Pi}
\renewcommand\S{\Sigma}
\newcommand\Th{\Theta}

\newcommand\G{\Gamma}
\newcommand\F{\Phi}

\newcommand{\lag}{\langle}
\newcommand{\rag}{\rangle}

\newcommand{\cA}{{\cal A}}

\newcommand{\cD} {{\mathcal D}}

\newcommand{\cO}{{\cal O}}
\newcommand{\cS}{{\cal S}}

\newcommand{\cT}{{\cal T}}

\newcommand{\pa}{\partial}

\newcommand{\nn}{\nonumber \\}
\newcommand{\na}{\nabla}

\newcommand{\sdfrac}[2]{\mbox{\small$\displaystyle\frac{\raisebox{-1.8pt}{${#1}$}}{\raisebox{1pt}{${#2}$}}$}}

\newcommand{\bea}{\vspace{-3mm}\begin{eqnarray}}
\newcommand{\eea}{\vspace{-3mm}\end{eqnarray}}
\newcommand{\be}{\vspace{-3mm}\begin{equation}}
\newcommand{\ee}{\vspace{-3mm}\end{equation}}
\newcommand{\beo}{\vspace{-1mm}\begin{equation}}
\newcommand{\eeo}{\vspace{-1mm}\end{equation}}
\newcommand{\bet}{\vspace{-2mm}\begin{equation}}
\newcommand{\eet}{\vspace{-2mm}\end{equation}}
\newcommand{\bes}{\begin{subequations}}
\newcommand{\ees}{\end{subequations}}

\def\nbox#1#2{\vcenter{\hrule \hbox{\vrule height#2in
\kern#1in \vrule} \hrule}}
\def\sq{\,\raise1pt\hbox{$\nbox{.10}{.10}$}\,}
\def\sqb{\,\raise.5pt\hbox{$\overline{\nbox{.09}{.09}}$}\,}

\newcommand{\lrpr}{\raise 1ex\hbox{$^\leftrightarrow$} \hspace{-9pt} \partial}

\newcommand{\lpr}{\raise 1ex\hbox{$^\leftarrow$} \hspace{-9pt}\partial}
\newcommand{\rpr}{\raise 1ex\hbox{$^\rightarrow$} \hspace{-9pt}\partial}

\usepackage[T1]{fontenc} 
\usepackage[titletoc]{appendix}
\usepackage{setspace}
\allowdisplaybreaks
\begin{document}
\pagestyle{plain}

\title{Gravitational Vacuum Polarization: Decoupling and the Conformal Anomaly}

\author{Emil Mottola}
\email{mottola.emil@gmail.com, emottola@unm.edu}
\affiliation{Department of Physics and Astronomy, University of New Mexico\\
Albuquerque NM 87131\\}

\begin{abstract}
A compact form of the stress tensor $\lag T^{ab}T^{cd}\rag$ correlation function of a quantum scalar field of arbitrary mass $m$
and curvature coupling $\x$ is presented, with particular emphasis on decoupling in the $m\!\to\!\infty$ limit and infrared origin of the 
conformal anomaly as $m\!\to\!0$. The Ward Identity (WI) of covariant conservation is verified for the full second order metric variation 
of the one-loop effective action, of which $\lag T^{ab}T^{cd}\rag$ is part, including local contact terms. This WI is satisfied by two tensors, 
one spin-2 (traceless) and the second spin-0 (traceful), each multiplied by a Lorentz invariant form factor computed in $n$-dimensional 
regularization. Minimal subtraction of pole terms at $n\!=\!4$ produces form factors that fail to satisfy decoupling for general $\x \!\neq\!1/6$, but 
can be simply amended to do so.  Particular interest attaches to the $\x \!=\!1/6$ case, in which the spin-0 form factor at $n\!=\!4$ 
is completely finite, requires no UV regularization or subtractions, and satisfies decoupling directly. Its imaginary part defines a spectral 
function that obeys a UV finite sum rule for any $m$, while its real part unambiguously determines the $\sq R$ term in the conformal anomaly 
in the massless limit. Its $m^2$ dependence describes a Wilsonian renormalization group flow to an effective field theory limit at large distances. 
At $m\!=\!0$ the spin-0 spectral function becomes a Dirac $\d$-function at zero energy, demonstrating the existence of a massless scalar 
Goldstone collective excitation due to the conformal anomaly in the low energy effective theory of gravity.

\end{abstract} 

\date{July 19, 2026}
\maketitle
\vfil
\eject

\pagenumbering{arabic}
\pagestyle{plain}
\vfil\break

\section{Introduction: Effective Field Theory and Anomalies}

Effective Field Theories \cite{Wilson:1975,Georgi:1993,Manohar:2017,EFTPet} are based on general symmetry arguments and decoupling of heavy degrees 
of freedom from low energy physics~\cite{AppCar:1975}. The Effective Field Theory (EFT) approach is especially useful when the ultraviolet (UV) complete 
theory is not known, as in the case of gravity. The question arises then of how the effects of quantum fields such as those of the Standard Model are 
to be taken systematically into account in a low energy EFT of gravity far below the Planck energy, including at macroscopic distance scales.

Quantum field theory (QFT) effects on and due to gravitational fields are contained in the one particle irreducible (1PI) effective action $\cS_{\rm 1PI}[g]$ 
of the QFT embedded in a curved spacetime. Although general features of the quantum 1PI effective action are known, and its UV divergences 
can be determined through an asymptotic expansion of the heat kernel~\cite{DeWitt:1975,Shap:2008,BuchShap}, the low energy or infrared (IR) 
limit of $\cS_{\rm 1PI}[g]$ for a general four dimensional spacetime metric $g_{ab}$ is not determined by this expansion. Infrared effects must be
investigated by other methods, such as through $n$-point correlation functions of the stress-energy tensor (SET).

The simplest such function is the 1-point function and renormalized expectation value of $\big\lag \hat T^{ab}(x)\big\rag\mqs_g$, upon which a significant
amount of effort was expended, since it can be regarded as the source for Einstein's eqs.~in semi-classical gravity~\cite{BirDav}. However, this is 
problematic for several reasons. By standard methods such as dimensional regularization,  $\big\lag \hat T^{ab}(x)\big\rag$ in flat space is found 
to depend on the fourth power of heavy masses, which conceivably could be as large as the Planck mass. This strong dependence on the highest masses 
of QFT leads directly to the `cosmological constant problem'~\cite{WeinbergRMP:1989, Martin:2012}, and `unnaturalness' of the value of $\La$ dark energy 
inferred from cosmology~\cite{DES:2019}.

Apart from this severe discrepancy with observations, such dependence of a supposedly measurable parameter of the low energy EFT on heavy masses
violates the assumption of decoupling of quantum loops of heavy mass fields that is central to the EFT approach. The renormalized 
$\big\lag \hat T^{ab}(x)\big\rag\mqs_g$, like the full 1PI action $\cS_{\rm 1PI}[g]$, is the result of integrating out QFT modes at all scales, 
and thus contains both UV and IR effects. Since it is evaluated at a single point, $\big\lag \hat T^{ab}(x)\big\rag\mqs_g$ provides no information 
on the scale dependence of quantum correlations or a separation of UV from IR effects, necessary for a Wilsonian EFT at large distances or low energies.

One of the main results of earlier studies of $\big\lag \hat T^{ab}(x)\big\rag\mqs_g$ that is relevant to the low energy EFT of gravity,
is the conformal anomaly due to massless fields~\cite{CappDuff:1974, Brown:1977, BrownCass:1977,Duff:1977}. The conformal anomaly 
is analogous to the chiral anomaly in the divergence of the axial current $\big\lag\pa_\m J^\m_5(x)\big\rag$ of fermions coupled to background 
gauge fields~\cite{Adler:1969}. Anomalies play a special role in EFTs since they modify the symmetries and Ward Identities of the theory and 
give rise to additional terms in the low energy effective action. The axial anomaly in QCD leads to verifiable predictions 
for the $\p^0 \!\to \!\g \g$ decay rate~\cite{Adler:1969,Bertlbook,TreiJackGross:2015},  EFT meson self-interactions~\cite{WZ:1971}, and the 
solution of the $U(1)$ problem~\cite{tHooft:1986}, accounting for the $\h$ or $\h'$ singlet meson mass. These low energy consequences 
of the axial anomaly in QCD are remarkable examples of  `anomaly matching,'  relating the high energy UV theory of quarks and gluons to the 
low energy EFT of entirely different meson degrees of freedom in the spontaneously broken phase~\cite{tHooft:1979}. 

The low energy properties of the chiral anomaly become apparent by examining the full 3-point $\big\lag J^\m_5(z)J^\a(x) J^\b(y)\big\rag$ triangle (AVV) amplitude, 
prior to taking its divergence, which is in fact completely UV finite and determined, once Lorentz and gauge symmetries and their Ward Identities are enforced.
It is instructive to retain the fermion mass, in order to exhibit decoupling in the $m\!\to\!\infty$ limit, and the appearance of a massless pole in the opposite 
$m\!\to\!0$ limit~\cite{GiaEM:2009}. The imaginary part of the AVV triangle satisfies a UV finite sum rule for any $m$ that becomes a Dirac $\d(s)$ in the 
$m\!\to\!0$ limit~\cite{DolZak:1971, Horejsi:1985, GiaEM:2009}. The corresponding real part exhibits a massless pseudoscalar $1/k^2$ pole, the propagator 
of a collective Goldstone collective excitation in the $e^+e^-$ intermediate state of the AVV triangle amplitude independent of UV physics~\cite{GiaEM:2009}. 
\!\footnote{In an extenson of terminology the massless pole associated with the axial anomaly may be called a `Goldstone boson,' although it arises 
from the anomaly itself rather than from spontaneous symmetry breaking, and therefore represents an extension of the Goldstone theorem to the related
but distinct phenomenon of anomaly symmetry breaking (ASM)~\cite{SSEMAxion:2024}.} In Ref.~\cite{GiaEM:2009} this analysis was extended to the 
$\big\lag \hat T^{\m\n}(z)J^\a(x) J^\b(y)\big\rag$ amplitude where the axial vector vertex is replaced by the SET of Dirac fermions, showing that a massless 
scalar state is also produced by the conformal anomaly in QED in the $m\to 0$ limit, supporting the IR nature of both anomalies~\cite{ArmCorRose:2009}.

That the conformal anomaly is relevant to low energy EFT is seen also by its non-local effective action, massless pole and macroscopic effects it
implies~\cite{MazEMWeyl:2001,EMVau:2006,EMEFT:2022}. Exhibiting this massless scalar pole explicitly in SET correlators is significantly 
more involved than in the case of the chiral anomaly with vector or axial vector currents. The $\lag TTT\rag$ three-point function 
has been calculated in the exact conformal limit by solution of the anomalous conformal Ward Identities~\cite{CorRoseSerEM2013,BzoMcFSken:2018,BzoMcFSken:2014}, 
and is in perfect agreement with the anomaly effective action and massless Goldstone pole in the spin-0 trace sector~\cite{TTTCFT:2019}.
However, the use of conformal Ward Identities precludes the inclusion of mass breaking of conformal invariance, necessary to analyze the decoupling 
limit of the amplitude for large mass.

Since one must turn to higher point correlation functions of the SET in order to distinguish UV from IR effects and verify decoupling of heavy mass states, 
it is worthwhile to re-examine the next simplest two-point correlation function $\big\lag\hat T^{ab}(x)\hat T^{cd}(y)\big\rag$ 
at distinct spacetime points $x\!\neq \!y$ in a flat space background. This correlation function is the non-local part of the full second variation of 
the 1PI effective action $\cS_2^{abcd}(x,y)$, which includes also local contact terms, and can be computed by standard $n$-dimensional regularization
for a scalar field of arbitrary mass $m$ and curvature coupling $\x$~\cite{AndMolEM:2003, GorShap:2003}. When defined so that decoupling holds
in the $m\!\to\!\infty$ limit, the opposite $m\!\to\!0$ limit describes the IR properties of the low energy EFT, where the conformal anomaly and  
Goldstone pole emerges.

Maintaining the correct symmetries under quantization is essential for anomalies and for the EFT method, so the Ward Identity (WI) that $\cS_2^{abcd}$ 
must satisfy by virtue of the covariant conservation of $T^{ab}$ is derived in Sec.~\ref{Sec:WI}. This WI is verified in dimensional regularization, once the 
correct local contact  terms in $\cS_2^{abcd}$ of Sec.~\ref{Sec:TTdimreg} are taken into account. This leads to a relatively simple form of the full 
$\cS_2^{abcd}$ in terms of two tensor projectors, one spin-2 and traceless, the second spin-0 and traceful, together with their Lorentz invariant 
form factors in Sec.~\ref{Sec:Proj}. In Sec.~\ref{Sec:MS} the $1/(n-4)$ poles in each form factor are identified and removed by the standard modified minimal 
$\overline{MS}$ scheme. These are clearly UV short distance subtractions since the poles are absorbed by renormalization of strictly local terms in the 
gravitational action up to dimension four.  Although neither form factor defined this way satisfies decoupling for arbitrary $\x\! \neq\! 1/6$, a 
simple modification of the $\overline{MS}$ scheme by an additional finite subtraction of local terms is made in Sec.~\ref{Sec:Phys} to ensure decoupling. 
The resulting form factors define response functions to external spin-2 and spin-0 external metric perturbations of momentum scale $k^2$, whose logarithmic 
derivative with respect to $k^2$ defines the running of corresponding couplings with physical momentum, or equivalently as the mass $m$ itself is varied.

A by-product of this analysis and decomposition into spin-2 and spin-0 amplitudes satisfying the WI is that the UV renormalization of the $\La$ 
term is simply that required by the one-point function tadpole $\big\lag \hat T^{ab}(x)\big\rag$ term, which has no $k^2$ dependence and hence 
does not `run' with physical momentum scale $k^2$~\cite{ShapSol:2009}. The Newtonian gravitational constant $G$ behaves similarly and
also does not run with physical momentum.

Having verified the conservation WI, and ensured decoupling in the limit $m \to \infty$, one can make contact with the Wilsonian approach to EFT~\cite{Wilson:1975},
by viewing $1/m$ as a smooth, Lorentz invariant {\it infrared} cutoff, the dependence upon which realizes the flow to the low energy EFT limit as $m\!\to\!0$. 
Since the spin-0 form factor for $\x\!=\!1/6$ has no UV pole divergence at $n\!=\!4$, and automatically satisfies decoupling, its dependence on $k^2$ and $m$ 
is immediately physical, and independent of UV subtractions. That the $m\! \to\! 0$ limit of this finite function gives the coefficient of the 
$\sq R$ term of the conformal anomaly, as shown in Sec.~\ref{Sec:Anom}, establishes this $\sq R$ term in the anomaly as a genuine IR effect, 
independent of UV regularization, renomalization or UV physics of any kind, which therefore should be retained as a marginal term in the 
low energy EFT of gravity.

As shown in Sec.~\ref{Sec:SpecSum}, the absorptive part of the spin-$0$ part of the $\lag TT\rag$ polarization for $\x \!=\!1/6$ also
obeys a UV finite sum rule for any $m$, just as that found for the axial anomaly and trace anomaly $\lag TJJ\rag$ correlator in~\cite{GiaEM:2009}. 
Since this finite sum rule is independent of any UV regulator or subtraction, it is clearly a consequence and necessary feature of the low energy EFT. 
Moreover, in the limit $m \to 0$, this imaginary part spectral function becomes a Dirac $\d$-function, which is the signature of a gapless
collective excitation in the intermediate state scalar $0^+$ channel. The corresponding finite residue of the $1/k^2$ pole in the real part of $\lag TT\rag$,
unambiguously determines the coefficient of the $\sq R$ term in the conformal anomaly. Appendix~\ref{App:2D} shows the explicit massless pole
in the case of two dimensions, and Appendix~\ref{App:Ferm} extends this analysis and results to the gravitational polarization functions of massive 
Dirac fermions in four dimensions.

Although exhibiting the Goldstone pole explicitly in a physical correlation function requires going to the 3-point function as in~\cite{TTTCFT:2019},
analogous to the AVV or  $\lag TJJ\rag$ triangle diagrams, the defining features of either the axial or conformal anomaly that make both relevant 
at low energies or large distances can already be seen in the two-point $\lag TT\rag$ correlator.  Sec.~\ref{Sec:EFT} contains a summary of these results
and conclusions for the EFT of low energy gravity.

\section{Effective Action and Conservation Ward Identities in Position Space}
\label{Sec:WI}

The one-particle-irreducible (1PI) quantum effective action $\cS_{\rm 1PI}$ is defined formally by the functional integral over the quantum
matter fields, denoted generically by $\F$ as
\be
Z[g]= \int \big[\cD \F\big] \exp \Big \{i {\rm S_{cl}}[\F,g]\Big\}= \exp \Big\{i\cS_{\rm 1PI}[g]\Big\} \,,\qquad \cS_{\rm 1PI}[g] = -i \ln Z[g]
\label{Zdef}
\ee
in a general but fixed background metric $g_{ab}(x)$, where ${\rm S_{cl}}[\F,g]$ is their classical action.
The classical stress-energy tensor (SET) of $\F$ is defined by
\be
T^{ab}[\F, g;x] = \frac{2\!\!}{\!\!\sqrt{-g}} \,\frac{\d\, {\rm S_{cl}}}{\d g_{ab}(x)} \equiv \hat T^{ab}(x)
\label{Tdef}
\ee
where the hat notation $\hat T^{ab}(x)$ in (\ref{Tdef}) serves to remind that the classical SET becomes an operator upon quantization of $\F$.
At the classical level the SET satisfies
\be
\na\mqs_b \hat T^{ab}(x) = 0
\label{Tcons}
\ee
with $\na\mqs_a$ the covariant derivative with respect to the background metric $g_{ab}(x)$. This covariant conservation law follows from Noether's theorem 
provided ${\rm S_{cl}}$ is invariant under general coordinate transformations and $\F$ satisfies the classical equations of motion,
\be
\frac{\d\, {\rm S_{cl}}}{\d \F(x)} = 0\,.
\label{Heis}
\ee
In the quantum theory defined by (\ref{Zdef}), (\ref{Heis}) is replaced by the Heisenberg eqs.~of motion of the field operator $\hat \F$,
and (\ref{Tcons}) continues to hold for the expectation value $\lag \hat T^{ab}(x)\big\rag\mqs_g$
\be
\big\lag \hat T^{ab}(x)\big\rag\mqs_g = \frac{1}{Z} \int \big[\cD \F\big]\, T^{ab}[\F, g;x]\,  \exp \Big \{i {\rm S_{cl}}[\F,g]\Big\}
= \frac{2\!\!}{\!\!\sqrt{-g}}\, \frac{\d\, \cS_{\rm 1PI}[g]}{\d g_{ab}(x)}
\label{T1def}
\eeo
which also satisfies the covariant conservation law 
\be
\na\mqs_b \,\big\lag \hat T^{ab}(x)\big\rag\mqs_g  = 0
\label{Tqcons}
\ee
provided the SET operator and functional integral (\ref{Zdef}) is defined so that the quantum 1PI  effective action $\cS_{\rm 1PI}[g]$ is
also invariant under general coordinate transformations.

The higher point SET correlation functions are defined by the series expansion
\vspace{-3mm}
\begin{align}
&\hspace{-1cm}\cS_{\rm 1PI}[\bar g+h] = \cS_{\rm 1PI}[\bar g]\  +  \label{Tayexp} \\
&\hspace{-1.5cm}\sum_{n=1}^{\infty} \frac{1}{2^n\,n!}\! \int\! d^4x_1\dots d^4x_n \sqrt{-\bar g(x_1)} \dots \sqrt{-\bar g(x_n)}\, 
\cS^{\m_1\n_1\dots \m_n\n_n}_n(x_1, \dots,x_n;\bar g)\, h_{\m_1\n_1}(x_1)\dots h_{\m_n\n_n}(x_n)  \nonumber 
\end{align}
of the 1PI quantum effective action around a specific metric background $\bar g_{ab}$, where
\be
\cS_n^{\m_1\n_1\dots\m_n\n_n}(x_1,\ldots,x_n;\bar g) \equiv \frac{2^n}{\sqrt{-\bar g(x_1)}\dots\sqrt{-\bar g(x_n)}}\ 
\frac{\d^{n}\cS_{\rm 1PI}[g]}{\d g_{\m_1\n_1}(x_1)\dots \d g_{\m_n\n_n}(x_n)}\bigg\vert_{g= \bar g}
\label{Tn}
\ee
are the covariant time-ordered $\cT^*\{\dots\}$ product of SET operators 
\be
\cS_n^{\m_1\n_1\dots\m_n\n_n}(x_1,\ldots,x_n;\bar g) = 
i^{n-1}\,\Big\lag \cT^* \big\{\hat T^{\m_1\n_1}(x_1)\dots \hat T^{\m_n\n_n}(x_n)\big\}\Big\rag_{\bar g}\,,\qquad x_i \neq x_j \quad {\rm for} \quad i \neq j\,.
\label{correfn}
\ee
at distinct spacetime points $x_i$. For purposes of this paper $\bar g_{ab}$ will be set equal to the Minkowski metric 
$\h_{ab} = {\rm diag}\,(-1,1,1,1)$, and the expansion (\ref{Tayexp}) will be around flat space, after the variations in (\ref{Tn}) have been 
taken. The two-point connected correlation function
\bet
\cS_2^{abcd}(x,y) = 4\  \frac{\d^{2}\cS_{\rm 1PI}[g]}{\d g_{ab}(x)\,\d g_{cd}(y)}\bigg\vert_{g= \h}
= i\, \big\lag \hat T^{ab}(x)\,\hat T^{cd}(y)\big\rag_{\! \rm con}\! \equiv \P^{abcd}(x,y) \quad {\rm for} \quad (x -y)^2 > 0 
\label{TT2}
\eet
can be defined for spacelike separations, where the time ordering is superfluous. The subscript $\bar g = \h$ will be dropped when the
flat space limit of the correlation function in the Minkowski vacuum is understood. 

\begin{figure}[h]
\vspace{-2mm}
\includegraphics[height=6cm,width=10cm, trim=0cm 0cm 0cm 0cm, clip]{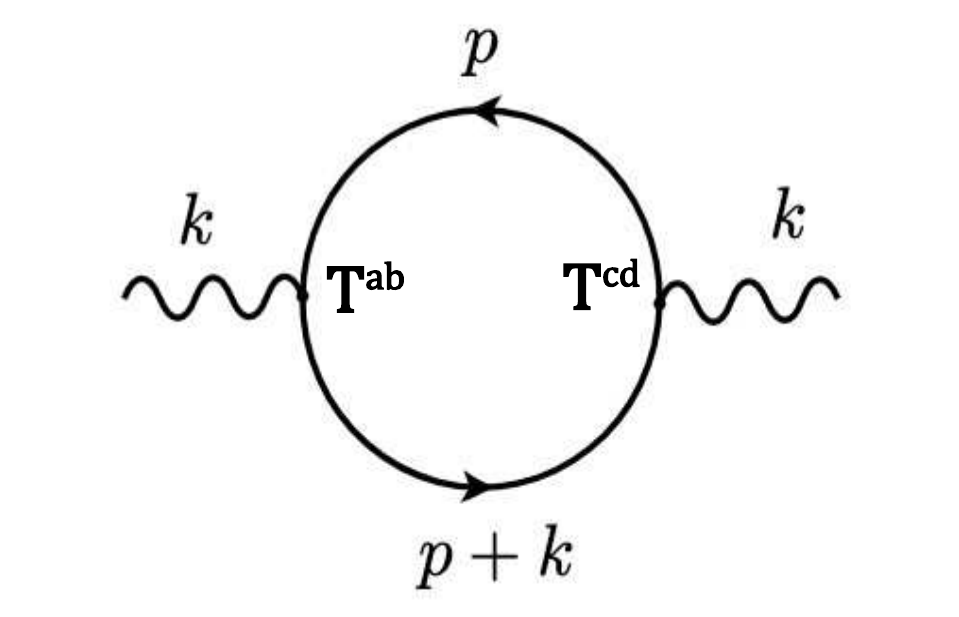}
\singlespace
\caption{The non-local gravitational vacuum polarization $\P^{abcd}(x,y)$ for $x\neq y$ in momentum space, where $k$ is the external $4$-momentum
flowing through the diagram, and $p$ is the internal loop $4$-momentum to be integrated over.}
\label{Fig:VacPol}
\vspace{-2mm}
\end{figure}

This correlation function measures the non-local `polarization' of the Minkowski vacuum at $x$ due to a gravitational metric perturbation 
$\d g_{cd} = h_{cd}(y)$ at $y$, {\it cf.} Fig.~\ref{Fig:VacPol}, and contributes to the graviton self-energy, analogous to the electromagnetic 
polarization tensor $\P^{ab}(x,y) = -i \, \big\lag \hat j^a(x)\,\hat j^b(y)\big\rag_{\! \rm con}$ due to an external gauge potential perturbation 
$\d A_b(y)$, which contributes to the photon self-energy. The sign of $\P^{abcd}(x,y)$ in (\ref{TT2}) is chosen so that its Fourier transform 
$\P^{abcd}(k)$ is in agreement with the conventions of Ref.~\cite{AndMolEM:2003}.\footnote{The opposite sign convention may be more 
appropriate in other contexts, since (\ref{TT2}) results in a negative graviton self-energy in momentum space, {\it cf.} (\ref{FkP}).}

When the spacetime points coincide at $x\!=\!y$, additional local terms proportional to Dirac $\d$-functions and derivatives thereof appear 
in $\cS_2^{abcd}$, modifying (\ref{TT2}) and requiring special care. From the definitions (\ref{Zdef})-(\ref{Tdef}), we have
\vspace{-2mm}
\begin{align}
\frac{\d^2\,\cS_{\rm 1PI}[g]}{\d g_{ab}(x)\,\d g_{cd}(y)} &= 
\frac{1}{Z}\, \bigg\lag \frac{\d^2\,{\rm S_{cl}}}{\d g_{ab}(x)\,\d g_{cd}(y)}\bigg\rag 
+ \frac{i}{Z} \, \bigg\lag \frac{\d\,{\rm S_{cl}}}{\d g_{ab}(x)}\, \frac{\d\,{\rm S_{cl}}}{\d g_{cd}(y)}\bigg\rag 
- \frac{i}{Z}\, \bigg\lag \frac{\d\,{\rm S_{cl}}}{\d g_{ab}(x)}\bigg\rag  \  
\frac{1}{Z}\,\bigg\lag \frac{\d\,{\rm S_{cl}}}{\d g_{cd}(y)}\bigg\rag \nn[4pt]
&= \frac{1}{2} \bigg\lag \frac{\d \big[\sqrt{-g}\, \hat T^{ab}(x)\big]}{\d g_{cd}(y)}\bigg\rag 
+ \frac{i}{4}\sqrt{-g(x)}\sqrt{-g(y)} \  \Big\lag\hat T^{ab}(x)\, \hat T^{cd}(y) \Big\rag_{\!\rm con}
\label{var2S}
\end{align} 
around a general background, where
\vspace{-2mm}
\be
\Big\lag\hat T^{ab}(x)\, \hat T^{cd}(y) \Big\rag_{\!\rm con} \!\!=  \Big\lag\hat T^{ab}(x)\, \hat T^{cd}(y) \Big\rag
- \Big\lag\hat T^{ab}(x)\Big\rag\, \Big\lag\hat T^{cd}(y) \Big\rag
\eeo
is the connected two-point $\big\lag TT\big\rag$ correlator. Thus (\ref{Tn}) evaluated for $\cS_2^{abcd}$ around flat space is
\be
\cS_2^{abcd}(x,y) =4\  \frac{\d^{2}\cS[g]}{\d g_{ab}(x)\,\d g_{cd}(y)}\bigg\vert_{g= \h}= 
 i\,  \Big\lag\hat T^{ab}(x)\, \hat T^{cd}(y) \Big\rag_{\!\rm con} +\ 2\ \bigg\lag \frac{\d \big[\sqrt{-g}\, \hat T^{ab}(x)\big]}{\d g_{cd}(y)}\bigg\rag_{\!\h} 
\label{S2tot}
\ee
showing the additional local contact term in $\cS_2^{abcd}(x,y)$ with support only at $x=y$, compared to (\ref{TT2}) 
and Fig.~\ref{Fig:VacPol} at unequal spacelike separated points.

The Ward Identities satisfied by the multi-point functions $\cS_n^{\m_1\n_1\dots\m_n\n_n}(x_1,\ldots,x_n)$ are obtained by taking 
successive metric variations of the basic general covariant conservation law (\ref{Tqcons}). Expressed in a convenient form, this is
\be
\sqrt{-g}\,\na\mqs_b\, \big\lag \hat T^{ab}(x)\big\rag\mqs_g = 
2\,\frac{\pa}{\pa x^b}\left(\frac{\d\cS_{\rm 1PI}[g]}{\d g_{ab}(x)}  \right)  
+ 2\,\G^a_{\ \,\m\n}(x) \left(\frac{\d \cS_{\rm 1PI}[g]}{\d g_{\m\n}(x)} \right)  = 0
\label{onept}
\ee
by making use of the definitions of (\ref{T1def}) and the Christoffel connection 
\be
\G^a_{\ \,\m\n}(x) = \sdfrac{1}{2}\, g^{a\a}\,\big(\! - \pa_\a g_{\m\n} +  \pa\mqs_\m g_{\n\a} +  \pa_\n g_{\a\m} \big)\,.
\label{Chris}
\ee
Since (\ref{onept}) is valid for an arbitrary metric, it may be varied by $\d g_{cd}(y)$ to obtain
\beo
2\,\frac{\pa}{\pa x^b}\left(\frac{\d^2\,\cS_{\rm 1PI}[g]}{\d g_{ab}(x)\,\d g_{cd}(y)}  \right)  
+ 2\,\frac{\d \G^a_{\ \,\m\n}(x)}{\d g_{cd}(y)} \left(\frac{\d \cS_{\rm 1PI}[g]}{\d g_{\m\n}(x)} \right)
+ 2\,\G^a_{\ \,\m\n}(x) \left(\frac{\d^2 \cS_{\rm 1PI}[g]}{\d g_{\m\n}(x)\,\d g_{cd}(y)} \right) = 0\,.
\label{varone}
\eeo
Evaluating this in flat space with Minkowski metric $\h^{ab}$ for which $\G^a_{\ \,\m\n}$ vanishes, and making use of (\ref{T1def}) and
(\ref{S2tot}), we have
\beo
\hspace{-4mm}\frac{\pa}{\pa x^b}\,\cS_2^{abcd}(x,y)  
+ 2\, \frac{\d \G^a_{\ \,\m\n}(x)}{\d g_{cd}(y)}\bigg\vert_{\h} \Big\lag \hat T^{\m\n}(x)\Big\rag_{\!\h} = 0 
\label{dS2}
\eeo
for the covariant conservation Ward Identity satisfied by $\cS_2$.

Now from (\ref{Chris}), the variational derivative of $2\, \G^a_{\ \,\m\n}(x)$ in the flat space limit is 
\bet
2\ \frac{\d \G^a_{\ \,\m\n}(x)}{\d g_{cd}(y)}\bigg\vert_{g=\eta} = \eta^{a\a}
\left\{\!- \d_{\m}^{\ (c} \d_{\ \n}^{d)}\,\frac{\pa}{\pa x^\a} + \d_{\n}^{\ (c}\d^{d)}_{\ \a} \,\frac{\pa} {\pa x^\m}
+ \d_{\a}^{\ (c}\d^{d)\!\!\!}_{\ \m}\, \frac{\pa}{\pa x^\n}\right\}\,\d^4(x - y)
 \label{varChris}
\eet
and since the one-point function must be proportional to the Minkowski metric tensor 
\be
\Big\lag \hat T^{\m\n}(x)\Big\rag_{\!\h} = K\, \h^{\m\n}
\label{Kdef}
\ee
by Lorentz invariance, with some constant $K$ in order to satisfy (\ref{Tqcons}), we find
\be
\frac{\pa}{\pa x^b}\, F^{abcd}(x,y) = 0
\label{WI}
\ee
where 
\vspace{-5mm}
\begin{align}
&F^{abcd}(x,y)  \equiv \cS_2^{abcd}(x,y) + K \,\big(\!-\h^{ab}\h^{cd} + \h^{ac}\h^{bd} + \h^{ad}\h^{bc}\big)\,\d^4(x-y)\nn[6pt]
& \hspace{-5mm}= i \,\Big\lag\hat T^{ab}(x)\, \hat T^{cd}(y) \Big\rag_{\!\rm con}
+  2\, \bigg\lag \frac{\d\, \big[\mqs\!\sqrt{-g}\, \hat T^{ab}(x)\big]}{\d g_{cd}(y)}\bigg\rag_{\!\h}
+ K\,\big(\!-\h^{ab}\h^{cd} + \h^{ac}\h^{bd} + \h^{ad}\h^{bc}\big)\,\d^4(x-y)
\label{Fdef}
\end{align}
which contains a second $\d$-function contact term, in addition to that in (\ref{S2tot}). This second contact term arises from the simple tadpole 
$\big\lag \hat T^{\m\n}(x)\big\rag_\h$ of (\ref{Kdef}).

\begin{figure}[h]
\vspace{-6mm}
\includegraphics[height=3cm,width=4.5cm, trim=0cm 0cm 0cm 0cm, clip]{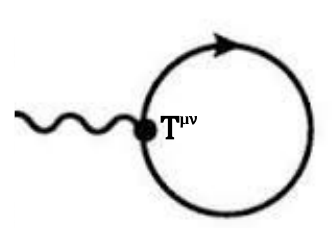}
\singlespace
\caption{The local tadpole term $\big\lag \hat T^{\m\n}(x)\big\rag$ of (\ref{Kdef}) for $x=y$ contributing to (\ref{Fdef}). }
\label{Fig:Tad}
\vspace{-2mm}
\end{figure}

Thus, when the two local contact terms in (\ref{Fdef}) with support only at coincident points $x\!=\!y$, are added to the non-local SET correlator
$\P^{abcd}(x,y)$ of (\ref{TT2}) for $x \!\neq\! y$ spacelike separated, the resulting function $F^{abcd}(x,y)$ satisfies the simple homogeneous 
covariant conservation Ward Identity (\ref{WI}) in flat Minkowski space. Note that whereas the non-local polarization tensor $\P^{abcd}(x,y)$ 
contains both real and imaginary parts, the two local contact terms in (\ref{Fdef}) are purely real. 

\section{Evaluation of (\ref{Fdef}) for a Scalar Field in Dimensional Regularization}
\label{Sec:TTdimreg}

In this section the general formalism of the previous section will be applied to the case of a free scalar field $\f$, with arbitrary mass $m$ and
curvature coupling $\x$, the classical action for which is 
\be
{\rm S_{cl}}[\f,g] = -\frac{1}{2}\int \! d^4 x  \sqrt{-g}\, \Big\{(\na\f)^2 +m^2 \f^2+ \x\, R\, \f^2\Big\}
\label{Sf} 
\ee
where $(\na\f)^2 = (\na\mqs_a\f)\, g^{ab}\, (\na\mqs_b \f)= (\na^a\f)\, g_{ab}\, (\na^b \f)$ and $R$ is the Ricci scalar. 
The SET of $\f$ in a general metric background, as defined by (\ref{Tdef}) is
\be
\hat T^{ab}(x) = \na^a \f \, \na^b \f  - \sdfrac{1}{2}\, g^{ab}\, \Big[  (\na\f)^2  + m^2 \f^2 \Big]
+ \x\, \Big[ G^{ab} + g^{ab} \sq - \na^a\na^b \Big]\, \f^2
\label{Tf}
\ee
where $G^{ab} = R^{ab} - g^{ab} R/2$ is the Einstein tensor.

\subsection{Non-local Vacuum Polarization}

Due to translation invariance of the vacuum state in flat Minkowski space, the non-local connected correlator 
$ i \,\big\lag\hat T^{ab}(x)\, \hat T^{cd}(y) \big\rag_{\!\rm con}$ at distinct spacelike separated spacetime points 
is a function only of their difference $x-y$. Its Fourier transform in $n$-dimensional flat space is therefore
\be
\P^{abcd}(k) =i\! \int\! d^n\!x \ e^{ik\cdot (x-y)}\ \Big\lag \hat T^{ab}(x)\, \hat T^{cd}(y)\Big\rag_{\!\rm con}
=  \int\! d^n\!x_E \ e^{ik\cdot (x-y)}\ \Big\lag \hat T^{ab}(x)\, \hat T^{cd}(y)\Big\rag_{\!\rm con}
\label{Pidef}
\ee
with the latter expression the Fourier transform in Euclidean time $d\t = i\,dt$. Because of Lorentz invariance of the vacuum state, 
the symmetries of $\P^{abcd}$ under interchange of $(ab)$ or $(cd)$, and the pair $(ab) \leftrightarrow (cd)$, the tensor $\P^{abcd}(k)$ 
must be expressible as the sum
\be
\P^{abcd}(k)  = \sum_{j=1}^{5} \P\mqs_j (k^2) \, \t_j^{abcd}(k)
\label{Pitau}
\ee
over the five basis tensors,
\vspace{-5mm}
\bes
\begin{align}
\t_1^{abcd}(k)&= \h^{ab}\h^{cd} \\
\t_2^{abcd}(k)&= \h^{ac}\h^{bd} +  \h^{ad}\h^{bc}\\
\t_3^{abcd}(k)&= \h^{ab} k^c k^d +\h^{cd} k^a k^b\\
\t_4^{abcd}(k)&= \h^{ac} k^b k^d + \h^{ad} k^b k^c + \h^{bc} k^a k^d + \h^{bd} k^a k^c\\
\t_5^{abcd}(k)&= k^a k^b k^c k^d
\end{align}\label{taudef}
\ees

\vspace{-1cm}
\noindent
with $\P_j (k^2), j= 1, \dots, 5$ Lorentz invariant scalar functions of $k^2, m, \x$ and  $n$.
The first two tensors $\t_{1,2}$ of (\ref{taudef}) are in fact independent of $k$.

The five functions $\P_j(k^2)$ were computed for the scalar field of (\ref{Sf})-(\ref{Tf}) in $n$-dimensional flat space in Refs.~\cite{AndMolEM:2003,GorShap:2003}.
The results of \cite{AndMolEM:2003} for the strictly non-local polarization parts are denoted here by $\P_j(k^2)$ since the notation
$F_j(k^2)$ is reserved for the full $F^{abcd}$ of (\ref{Fdef}) in (\ref{Fitot}) including local terms. After some algebraic simplifications,
the results of \cite{AndMolEM:2003} can be expressed as
\vspace{-3mm}
\bes
\begin{align}
&\P_1(k^2)\big\vert_{\x = 0} \!= (4\p)^{\!-\medmath{\sdfrac{n}{2}}}\, 
 \G\left(2 - \sdfrac{n}{2}\right)\! \int_0^1\! dx\,  \big(M^2\big)^{\medmath{\sdfrac{n}{2}\!-\!2}}\, \Big[\, 2 k^4 x^2 (1-x)^2 \Big]\nn
&\hspace{2.2cm} +  (4\p)^{\!-\medmath{\sdfrac{n}{2}}}\,  \G\left(- \sdfrac{n}{2}\right)\! \int_0^1\! dx\,\left[ -\sdfrac{1}{2}\,\big(M^2\big)^{\medmath{\sdfrac{n}{2}}} + \sdfrac{1}{4}\, n\,k^2x\,(1-2x) \big(M^2\big)^{\medmath{\sdfrac{n}{2}\!-\!1}}\right]
\label{F10}\\[2pt]
&\P_2(k^2)\big\vert_{\x = 0} =  (4\p)^{-\medmath{\sdfrac{n}{2}}}\,  \G\left(- \sdfrac{n}{2}\right) \int_0^1 dx\ 
\left[\sdfrac{1}{2}\, \big(M^2\big)^{\medmath{\sdfrac{n}{2}}}\right]\\[2pt]
&\P_3(k^2)\big\vert_{\x = 0} = (4\p)^{\!-\medmath{\sdfrac{n}{2}}}\, 
 \G\left(2 - \sdfrac{n}{2}\right) \int_0^1 dx\  \big(M^2\big)^{\medmath{\sdfrac{n}{2}\!-\!2}}\, 
\Big[\!-\! 2 k^2 x^2 (1-x)^2\Big]\nn[2pt]
&\hspace{2.2cm} + (4\p)^{\!-\medmath{\sdfrac{n}{2}}}\,  \G\left(1 - \sdfrac{n}{2}\right) 
\int_0^1 dx\  \big(M^2\big)^{\medmath{\sdfrac{n}{2}\!-\!1}}\ \Big[x\, (1-2x)\Big] \\
&\P_4(k^2)\big\vert_{\x = 0}= (4\p)^{\!-\medmath{\sdfrac{n}{2}}}\, 
 \G\left(1 - \sdfrac{n}{2}\right) \int_0^1 dx\  \big(M^2\big)^{\medmath{\sdfrac{n}{2}\!-\!1}} \left[-\sdfrac{1}{2}\, x\, (1-2x) \right]\\[2pt]
&\P_5(k^2)\big\vert_{\x = 0}= (4\p)^{\!-\medmath{\sdfrac{n}{2}}}\, 
 \G\left(2 - \sdfrac{n}{2}\right) \int_0^1 dx\  \big(M^2\big)^{\medmath{\sdfrac{n}{2}\!-\!2}} \left[2 x^2 (1-x)^2\right]
\end{align} \label{Fi}
\ees
for the $\x = 0$ part of (\ref{Pidef}), where 
\be
k^4 \equiv (k^2)^2 \,, \qquad M^2 \equiv k^2 x(1-x) + m^2
\label{M2def}
\ee 
and the integrations in (\ref{Fi}) are over the Feynman parameter $x \in [0,1]$.

The additional $\x$ dependent parts of (\ref{Pidef}) are given by
\vspace{-3mm}
\bes
\begin{align}
&\P_1 (k^2)\big\vert_{\x\,{\rm terms}}\!= (4\p)^{\!-\medmath{\sdfrac{n}{2}}}\, 
 \G\left(2 - \sdfrac{n}{2}\right)\! \int_0^1\! dx\,  \big(M^2\big)^{\medmath{\sdfrac{n}{2}\!-\!2}}\  \big(2\x k^4\big) \Big[\x - 2 x(1-x) \Big]\\
&\P_3(k^2)\big\vert_{\x\,{\rm terms}}= (4\p)^{\!-\medmath{\sdfrac{n}{2}}}\, 
 \G\left(2 - \sdfrac{n}{2}\right) \int_0^1 dx\  \big(M^2\big)^{\medmath{\sdfrac{n}{2}\!-\!2}}\,  \big(\!-\!2\x k^2\big)\Big[\x - 2 x(1-x) \Big]\\
&\P_5(k^2)\big\vert_{\x\,{\rm terms}}= (4\p)^{\!-\medmath{\sdfrac{n}{2}}}\, 
 \G\left(2 - \sdfrac{n}{2}\right) \int_0^1 dx\  \big(M^2\big)^{\medmath{\sdfrac{n}{2}\!-\!2}}\  \big(2\x\big) \Big[\x - 2 x(1-x) \Big]
\end{align} \label{Fixi}
\ees
with
\be
\P_2(k^2)\big\vert_{\x\,{\rm terms}}= \P_4 (k^2)\big\vert_{\x\,{\rm terms}}= 0\,. 
\label{F2F4xi}
\eeo
Defining the projection operator onto transverse vectors
\be
\th^{ab} = \h^{ab} - \frac{k^a\,k^b}{k^2}
\label{thdef}
\ee
so that
\be
k^4\, \th^{ab}\,\th^{cd} =  k^4\t_1^{abcd} - k^2\t_3^{abcd} + \t_5^{abcd}
\eeo
the $\x$-dependent part of (\ref{Pitau}) may be written
\be
\P^{abcd}(k)\big\vert_{\x\,{\rm terms}} = \Big[k^4 \,\th^{ab}\,\th^{cd}\Big] \,(4\p)^{\!-\medmath{\sdfrac{n}{2}}}\, 
 \G\left(2 - \sdfrac{n}{2}\right)\!  \int_0^1\! dx\,  \big(M^2\big)^{\medmath{\sdfrac{n}{2}\!-\!2}}\ 
\Big[2\,\x^2 - 4\,\x x(1-x) \Big]\,.
\label{Pixi}
\ee

\subsection{Local Contact Terms}
\label{Sec:local}

For computing the local contact terms at $x=y$ in (\ref{S2tot}) and (\ref{Fdef}) we need 
\be
\big\lag \f^2 \big \rag = \int\! \frac{d^n\mqs p}{(2 \p)^n}\frac{1}{p^2 + m^2} = (4\p)^{\!-\medmath{\sdfrac{n}{2}}}\, 
 \G\left(1 - \sdfrac{n}{2}\right)\, m^{\medmath{n\!-\!2}}
\label{phisq}
\ee
where the relation $\big\lag \f^2(x) \big\rag = G_E (x,x)$ for the coincident limit of the Euclidean Green's function $G_E(x,y)$ for $y=x$, expressed as the 
$n$-dimensional Euclidean momentum Fourier transform over $p\!=\!p_E$ evaluated at Euclidean time $\t = it$ has been used. Eq.~(\ref{phisq}) is valid for $n\! < \!2$, and 
analytically continued to complex $n$. Then since the $\G$ function satisfies $z \, \G(z) = \G(1 + z)$, 
\be
\big\lag \hat T^{ab}\big\rag = -\,\frac{m^2\!}{\!n}\,\big\lag \f^2 \big \rag \ \h^{ab} 
= (4\p)^{\!-\medmath{\sdfrac{n}{2}}}\, \G\left(\! - \sdfrac{n}{2}\right)\, \frac{m^n}{\!2}\ \h^{ab} 
\label{S1T}
\eeo
which determines the constant $K$ in (\ref{Kdef}) and (\ref{Fdef}) to be
\be
K = -\,\frac{m^2}{\!n}\,\big\lag \f^2 \big \rag =  (4\p)^{\!-\medmath{\sdfrac{n}{2}}}\, \G\left(\! - \sdfrac{n}{2}\right)\, \frac{m^n}{\!2}
\label{Kval}
\eeo
for the scalar $\f$ field in dimensional regularization. As usual, an arbitrary mass scale $\m$ will be introduced to define $m^n$ and $m^{n-2}$ 
and preserve dimensions away from $n=4$. 

From (\ref{Tf}) the variation in the contact term of (\ref{S2tot}) is
\vspace{-3mm}
\begin{align}
&\d\, \big[\!\!\sqrt{-g}\, \hat T^{ab}(x)\big]= \d \, \big[\!\!\sqrt{-g}\, g^{a\m}g^{b\n}\big] \, \hat T_{\m\n}(x)
+ \big[\!\!\sqrt{-g}\, g^{a\m}g^{b\n}\big] \, \d\hat T_{\m\n}(x)\nn
&\hspace{1cm}=\sqrt{-g}\,\left\{\sdfrac{1}{2} g^{cd} \,\hat T^{ab} - g^{ac} \,\hat T^{bd} - g^{bc} \,\hat T^{ad}\right\} \d g_{cd}(x)
+ \sqrt{-g} \, g^{a\m}g^{b\n} \d \hat T_{\m\n}(x)
\end{align}

\vspace{-7mm}
\noindent
where
\be
\d \hat T_{\m\n} (x) = \sdfrac{1}{2} \Big\{ -  \big[\big(\na \f\big)^2 + m^2 \f^2 \big] \, g^{ac}g^{bd} 
+ g^{ab}\, \na^c\mqs\f\,  \na^d\mqs\f\Big\}\, \d g_{cd}(x) + \x\, \d\,  \Big[ G^{ab} + g^{ab} \sq - \na^a\na^b \Big]\, \f^2\,.
\label{varTmn}
\ee
Evaluating the expectation value of this variation in dimensional regularization gives for the $\x =0$ terms
\begin{align}
2\, \bigg\lag \frac{\d\, \big[\mqs\!\sqrt{-g}\, \hat T^{ab}(x)\big]}{\d g_{cd}(y)}\bigg\rag_{\!\h\,, \x=0} &= 
\left\{K\, \h^{ab}\h^{cd}- 2K\, \h^{ac}\h^{bd} - 2K\, \h^{bc}\h^{ad} + K\, \h^{ab} \h^{cd}\right\}\,\d^n\mqs (x-y)\nn
&=2\,K\, \big(\t_1^{abcd}- \t_2^{abcd}\big)\,\d^n\mqs (x-y)
\label{varTloc}
\end{align}

\vspace{-7mm}
\noindent
in the flat space limit, where
\vspace{-2mm}
\begin{align}
&\big\lag\na^c\mqs\f\,  \na^d\mqs\f\big\rag =  \int\! \frac{d^n\mqs p}{(2 \p)^n}\frac{p^c p^d}{p^2 + m^2}
= - \h^{cd}\, (4\p)^{\!-\medmath{\sdfrac{n}{2}}}\, \G\left(1 - \sdfrac{n}{2}\right)\, \frac{m^n}{n}
= - \h^{cd}\, \frac{m^2\!}{\!n}\big\lag \f^2 \big \rag = K\h^{cd}\\[6pt]
& \hspace{1cm} {\rm and} \qquad  \Big\lag \big(\na \f\big)^2 + m^2 \f^2 \Big\rag 
= \h_{cd} \,\Big\lag\na^c\mqs\f\,  \na^d\mqs\f\Big\rag + m^2 \, \big\lag \f^2 \big \rag = 0
\end{align}

\vspace{-3mm}
\noindent
have been used. This combines with the last contact term in (\ref{Fdef}), which is exactly $1/2$ of (\ref{varTloc}) and of the opposite sign to give
\beo
F^{abcd}\big\vert_{{\rm loc},\, \x =0} = K\, \big(\t_1^{abcd}- \t_2^{abcd}\big)\,\d^n\mqs (x-y)
\label{Floc0}
\eeo 
for the sum of the two local terms in (\ref{Fdef}) for the scalar field in the flat space limit at $\x \!=\!0$. 

For the $\x\!\neq\! 0$ local terms in (\ref{Fdef}) from (\ref{varTmn}), the variation of the Einstein tensor $G^{ab}$ is  
\be
\d G^{ab}(x)\big\vert_\h = \sdfrac{1}{2}\,  \Big\{\h^{ab}\h^{cd}\sq - \h^{ac}\h^{bd}\sq -  \h^{ab} \pa^c\pa^d - \h^{cd} \pa^a\pa^b
+ \h^{ac} \pa^b \pa^d + \h^{bc}\pa^a\pa^d\Big\}\,  \d g_{cd} (x)
\ee
when evaluated in the flat space limit, so that this variation in (\ref{varTmn}) produces the $\x$-dependent local term
\begin{align}
2\x\,\left[\frac{\d G^{ab}(x)}{\d g_{cd}(y)}\big\lag \f^2\big\rag\right]_\h
&=\x\ \big\lag \f^2\big\rag\, \Big\{\h^{ab}\h^{cd}\sq -  \h^{a(c}\h^{d)b}\sq-  \h^{ab} \pa^c\pa^d - \h^{cd} \pa^a\pa^b
+  \h^{a(c} \pa^{d)} \pa^b + \h^{b(c}\pa^{d)}\pa^a\Big\}\,  \d^n (x-y)\nn
& \to \x\, \big\lag \f^2\big\rag\, \Big\{\!- k^2 \t_1^{abcd} + \sdfrac{1}{2}k^2 \t_2^{abcd} + \t_3^{abcd} - \sdfrac{1}{2} \t_4^{abcd}\Big\}
\label{Flocxi}
\end{align}
in momentum space, to be added to the variation (\ref{Floc0}) at $\x\!=\!0$.

Thus, (\ref{Floc0}) and (\ref{Flocxi}) are the total local contact terms in (\ref{Fdef}) to be added to the non-local 
polarization function $\P^{abcd}(k)$ given by (\ref{Pidef})-(\ref{Fi}). These local additions give for total $F^{abcd}$
\be
F^{abcd}(k) = \ \sum_{j=1}^{5} F\mqs_j\, (k^2) \, \t_j^{abcd}(k)
\label{Fk}
\ee
in the tensor basis (\ref{taudef}) in momentum space, with
\vspace{-3mm}
\bes
\begin{align}
F\mqs_1 (k^2) &= \P_1(k^2)\big\vert_{\x =0 }  + \P_1(k^2)\big\vert_{\x\,{\rm terms}}  + K - \x\, k^2 \big\lag \f^2\big\rag\\
F\mqs_2 (k^2) &= \P_2(k^2)\big\vert_{\x =0 }   -  K + \sdfrac{1}{2}\, \x\,  k^2 \big\lag \f^2\big\rag\\
F\mqs_3 (k^2) &= \P_3(k^2)\big\vert_{\x =0 }  + \P_3(k^2)\big\vert_{\x\,{\rm terms}} \ + \x \,\big\lag \f^2\big\rag\\
F\mqs_4 (k^2) &= \P_4(k^2)\big\vert_{\x =0 } -  \sdfrac{1}{2}\, \x \,\big\lag \f^2\big\rag\\
F\mqs_5 (k^2) &= \P_5(k^2)\big\vert_{\x =0 }  + \P_5(k^2)\big\vert_{\x\,{\rm terms}} \,.
\end{align}\label{Fisum}
\ees

\vspace{-1cm}
\noindent
Substituting in (\ref{Fisum}) the expressions for (\ref{Fi}) and (\ref{Fixi}), the resulting five form-factor scalar function $F_j (k^2)$ 
can be simplified somewhat by use of the identities
\be
\frac{\pa}{\pa x} (M^2)^\l= \l\, (M^2)^{\l-1}  \frac{\pa}{\pa x} (M^2) = \l\,  k^2 (1-2x) \, (M^2)^{\l-1} 
\label{dM2}
\ee
and consequentially
\be
\l \int_0^1 dx\,  k^2 x (1-2x) \, (M^2)^{\l-1} = x  (M^2)^\l\,\Big\vert_{x=0}^{x=1} - \int_0^1 dx\,  (M^2)^\l = 
m^{2\l} - \int_0^1 dx\,  (M^2)^\l
\label{intM}
\ee
following from (\ref{M2def}). Applying (\ref{intM}) to the last term in (\ref{F10}) for $\l = n/2$, and making use 
of (\ref{phisq}), (\ref{Kval}), and (\ref{Fisum}), we obtain 
\vspace{-3mm}
\bes
\begin{align}
&F_1(k^2)= (4\p)^{\!-\medmath{\sdfrac{n}{2}}}\, 
 \G\left(2 - \sdfrac{n}{2}\right)\! \int_0^1\! dx\,  \big(M^2\big)^{\medmath{\sdfrac{n}{2}\!-\!2}}\, 
(2 k^4) \Big[x^2 (1-x)^2 - 2\x x(1-x) + \x^2 \Big]\nn
&\hspace{2.2cm} +  (4\p)^{\!-\medmath{\sdfrac{n}{2}}}\,  \G\left(- \sdfrac{n}{2}\right)\! \int_0^1\! dx\,
\left\{ m^n -\big(M^2\big)^{\medmath{\sdfrac{n}{2}}} + \sdfrac{1}{2}\, n\, \x\,  k^2 m^{n-2} \right\} 
\label{F1}\\[2pt]
&F_2(k^2) =  (4\p)^{-\medmath{\sdfrac{n}{2}}}\,  \G\left(- \sdfrac{n}{2}\right) \int_0^1 dx\ 
\left\{\sdfrac{1}{2} \,\big(M^2\big)^{\medmath{\sdfrac{n}{2}}}\!\!- \sdfrac{1}{2}m^n -  \sdfrac{1}{4}\, n\,\x\,  k^2 m^{n-2} \right\}
\label{F2}\\[2pt]
&F_3(k^2) = (4\p)^{\!-\medmath{\sdfrac{n}{2}}}\, 
 \G\left(2 - \sdfrac{n}{2}\right)\! \int_0^1\! dx\,  \big(M^2\big)^{\medmath{\sdfrac{n}{2}\!-\!2}}\, 
(-2 k^2) \Big[x^2 (1-x)^2 - 2\x x(1-x) + \x^2 \Big]\nn
&\hspace{2.2cm} +  (4\p)^{\!-\medmath{\sdfrac{n}{2}}}\,  \G\left(1- \sdfrac{n}{2}\right)\! \int_0^1\! dx\,
\left\{\big(M^2\big)^{\medmath{\sdfrac{n}{2}}-1}x(1-2x) + \x\, m^{n-2} \right\} \label{F3}\\[2pt]
&F_4(k^2)= (4\p)^{\!-\medmath{\sdfrac{n}{2}}}\, \G\left(- \sdfrac{n}{2}\right) \int_0^1 dx\ 
\left\{\sdfrac{n}{4}\, \big(M^2\big)^{\medmath{\sdfrac{n}{2}\!-\!1}} x\, (1-2x) +\sdfrac{1}{4}\,n\, \x \, m^{n-2}\right\}
\label{F4}\\[2pt]
&F_5(k^2)= (4\p)^{\!-\medmath{\sdfrac{n}{2}}}\, \G\left(2 - \sdfrac{n}{2}\right) \int_0^1 dx\,
\big(M^2\big)^{\medmath{\sdfrac{n}{2}\!-\!2}}\, 2\, \Big[x^2 (1-x)^2 - 2\x x(1-x) + \x^2 \Big]\,. \label{F5}
\end{align}\label{Fitot}
\ees
\noindent
for the total $F_j$ in (\ref{Fk}) including all non-local and local contact terms.

\subsection{Conservation Ward Identities in Momentum Space}

Since the contractions
\vspace{-6mm}
\bes
\begin{align}
k_b\t_1^{abcd}(k)&=k^a\h^{cd} \\
k_b\t_2^{abcd}(k)&= k^c\h^{ad} +  k^d\h^{ac}\\
k_b\t_3^{abcd}(k)&= k^a k^c k^d + k^a\h^{cd} \ k^2\\
k_b\t_4^{abcd}(k)&= 2\, k^c k^a k^d + \big( k^c\h^{ad} +  k^d\h^{ac}\big)\, k^2\\
k_b\t_5^{abcd}(k)&= k^a k^c k^d\ k^2
\end{align}\label{tauk}
\ees

\vspace{-1cm}
\noindent
produce $3$ linearly independent tensors $k^a\h^{cd}, k^c\h^{ad} +  k^d\h^{ac}$, and $k^a k^c k^d$, setting each of their coefficients to zero in
\beo
k_b\, F^{abcd}(k) =  \sum_{j=1}^{5} F\mqs_j\, (k^2) \, k_b \t_j^{abcd}(k) = 0
\label{WIk}
\eeo
results in three relations among the five Lorentz invariant scalar form factors $F\mqs_j (k^2)$, namely
\vspace{-3mm}
\bes
\begin{align}
F_1 + k^2 F_3 &= 0\\
F_2 + k^2 F_4 &= 0\\
F_3 + 2F_4 + k^2 F_5 &=0\label{WIF3}\,.
\end{align}\label{WIF}
\ees

\vspace{-1cm}
Verifying that these relations hold for the dimensionally regulated $F_i$ of (\ref{Fitot}) is straightforward. The third of relations (\ref{WIF3}) 
follows immediately from (\ref{Fitot}), while the first two of (\ref{WIF}) are proven by making use of (\ref{intM}) for $\l= n/2$ in $k^2 F_3$ 
and $k^2 F_4$. Note that the contact terms (\ref{Floc0}) and (\ref{Flocxi}) are necessary to verify (\ref{WIF}), which consistent with 
(\ref{WI})-(\ref{Fdef}), do not hold for the non-local polarization terms $\P^{abcd}$ of (\ref{Pidef})-(\ref{Pixi}) alone. Since the local 
contact terms (\ref{Floc0})-(\ref{Flocxi}) are purely real, the imaginary parts of (\ref{WIF}) obtained only from the non-local terms of $\P^{abcd}$ 
do separately satisfy the WI, and that this is indeed the case was verified in~\cite{AndMolEM:2003}. Inclusion of the real local contact terms 
here enables the check of the WI (\ref{WIk}) and (\ref{WIF}) for the real parts as well.

\section{Spin-0, Spin-2 Projectors and Form Factors in $n=4$ Dimensions}
\label{Sec:Proj}

Having verified (\ref{WIF}), one may solve for $F_3, F_4$ and $F_5$ in terms of the two remaining independent form factors 
$F_1, F_2$, so that
\be
F^{abcd}(k) = F_1(k^2) \left( \t_1^{abcd} - \sdfrac{1}{k^2} \t_3^{abcd} + \sdfrac{1}{k^4}\t_5^{abcd}\right) 
+ F_2(k^2) \left(\t_2^{abcd}- \sdfrac{1}{k^2} \t_4^{abcd} + \sdfrac{2}{k^4} \t_5^{abcd}\right)\,.
\label{S2F1F2}
\eeo
It is convenient to introduce the scalar (spin-0 traceful) and tensor (spin-2 tracefree) projectors
\vspace{-2mm}
\bes
\begin{align}
P^{(S)\,abcd}(k) &= \sdfrac{1}{3}\, \th^{ab}\th^{cd}= \sdfrac{1}{3}\, \left( \t_1^{abcd} - \sdfrac{1}{k^2} \t_3^{abcd}
+ \sdfrac{1}{k^4}\t_5^{abcd}\right)\\[2pt]
P^{(T)\,abcd}(k) &= \sdfrac{1}{2}\,\Big(\th^{ac}\th^{bd} + \th^{ad}\th^{bc}\Big) - \sdfrac{1}{3}\, \th^{ab}\th^{cd}\nn
&\hspace{-1cm}=\frac{1}{2} \left(\t_2^{abcd}- \sdfrac{1}{k^2} \t_4^{abcd} + \sdfrac{2}{k^4} \t_5^{abcd}\right)
- \sdfrac{1}{3}\, \left( \t_1^{abcd} - \sdfrac{1}{k^2} \t_3^{abcd} + \sdfrac{1}{k^4}\t_5^{abcd}\right)
\end{align}
\label{PST}
\ees

\vspace{-1cm}
\noindent
in $n\!=\!4$ dimensions, with $\th^{ab}$ given by (\ref{thdef}), both of which are transverse under contraction with $k_b$. These 
projectors satisfy
\vspace{-5mm}
\bes
\begin{align}
P^{(A)\,ab\m\n}(k) \, P_{\ \ \, \m\n}^{(B)\ \ \,cd}(k)  &= P^{(A)\,abcd}(k) \ \d^{AB} \\
k_b P^{(A)\,abcd}(k) =0 &= P^{(A)\,abcd}(k)\,k_c\\
\h_{ab} P^{(T)\,abcd}(k) &= 0\\
\h_{ab} P^{(S)\,abcd}(k) &= \th^{cd} \label{thPS}
\end{align}\label{projs}
\ees
\vspace{-1.1cm}

\noindent
for $A, B = S, T$. Then (\ref{S2F1F2}) may be expressed in terms of these orthonormal projection tensors as
\be
F^{abcd}(k) = - \, \S(k^2)\,   P^{(S)\,abcd}(k) -\T(k^2)\, P^{(T)\,abcd}(k) 
\label{FkP}
\ee
in Fourier space, with 
\vspace{-3mm}
\begin{align}
&\S(k^2) =-3 F_1(k^2) - 2F_2 (k^2)\nn
&\hspace{2mm} = -(4\p)^{\!-\medmath{\sdfrac{n}{2}}}\,\G\left(2 - \sdfrac{n}{2}\right)\!
\int_0^1\! dx\,  \big(M^2\big)^{\medmath{\sdfrac{n}{2}\!-\!2}}\ \big(6\,k^4\big)\, \Big[x^2 (1-x)^2 - 2\x x(1-x) + \x^2 \Big]\nn
&\hspace{1cm}+ 2\, (4\p)^{-\medmath{\sdfrac{n}{2}}}\,  \G\left(- \sdfrac{n}{2}\right) \int_0^1 dx\ 
\left\{\big(M^2\big)^{\medmath{\sdfrac{n}{2}}} - m^n -  \sdfrac{1}{2}\, n\,\x\,  k^2 m^{n-2} \right\}
\label{Sigk}
\end{align}
the spin-0 traceful form factor, and
\be
\T (k^2)= -2F_2 (k^2) = -(4\p)^{-\medmath{\sdfrac{n}{2}}}\,  \G\left(- \sdfrac{n}{2}\right) \int_0^1 dx\ 
\left\{\big(M^2\big)^{\medmath{\sdfrac{n}{2}}}- m^n -  \sdfrac{1}{2}\, n\,\x\,  k^2 m^{n-2} \right\}
\label{Tk}
\ee
the spin-2 polarization tracefree form factor.  

The unrenormalized form factors (\ref{Sigk})-(\ref{Tk}) are defined by dimensional continuation away from $n\!=\! 4$, in order to identify and remove 
$1/(n-4)$ divergent pole terms, but the orthogonal projectors (\ref{PST}) are defined in the physical dimension $n\!=\!4$, as are the finite renormalized 
form factors after removal of the pole terms in Sec.~\ref{Sec:MS} below. The overall negative signs in (\ref{FkP}) are introduced in order to 
result in positive renormalized self-energy functions $\S_R(k^2), \T_R(k^2)$ for spacelike or Euclidean $k^2 > 0$, when evaluated at $n\!=\!4$ 
in (\ref{SRTR}) of the next section.

Eqs.~(\ref{FkP})-(\ref{Tk}), with (\ref{Fdef}) is a convenient form satisfying the WI (\ref{WIk}) for the subsequent analysis of the one-loop gravitational 
vacuum polarization of a quantum scalar $\f$ field with arbitrary mass $m$ and curvature coupling $\x$. The imaginary parts of  $\S(k^2), \T(k^2)$ 
agree with those previously given in \cite{AndMolEM:2003} by computing only the non-local terms of $\P^{abcd}$, and neglect of the contact terms 
(\ref{Floc0})-(\ref{Flocxi}), which are purely real.

\section{Identification of $1/(n-4)$ Poles and Renormalization in $\overline{MS}$\,Scheme}
\label{Sec:MS}

To facilitate the continuation to $n = 4$ and identification of $1/(n-4)$ poles, introduce the notation
\be
n=4-2 \e\,,\quad \e = 2 - \sdfrac{n}{2} \,,\quad - \sdfrac{n}{2} = \e -2
\label{epsdef}
\ee
in order to express
\vspace{-8mm}
\bes
\begin{align}
\S(k^2) &= \frac{6k^4\!}{(4 \p)^2}\, A -\frac{1}{(4\p)^2}\, B\\[2pt]
\T(k^2) & = \frac{1}{(4\p)^2}\, \frac{B}{2}
\end{align}
\label{STAB}
\ees
\vspace{-1cm}

\noindent
in terms of the two dimensionally regulated integrals
\vspace{-1mm}
\bes
\begin{align}
A&\equiv - (4\p)^\e \,\G(\e) \int_0^1\! dx\,  \left(\frac{M^2}{\m^2}\right)^{\!-\e} \Big[x^2 (1-x)^2 - 2\x x(1-x) + \x^2 \Big]\\[3pt]
B&\equiv  -2\, (4\p)^\e\,  \G(\e -2) \!\int_0^1\! dx\, \left\{\left(\frac{M^2}{\m^2}\right)^{\!-\e}\!\big(M^2\big)^2
-  \left(\frac{m^2}{\m^2}\right)^{\!-\e} \Big[ \big(2 -\e\big)\, \x k^2 m^2 + m^4\Big]\right\}
\end{align}
\label{ABdef}
\ees

\vspace{-1cm}
\noindent
where the arbitrary mass scale $\m$ has been introduced to preserve dimensionality. Now as $\e \to 0$
\vspace{-3mm}
\begin{align}
(4\p)^\e \,\G(\e)  &\to   \sdfrac{1}{\e} - \g_E + \ln (4 \p) + \cO(\e) = \sdfrac{1}{\bar \e} + \cO(\e)\nn
2\, (4\p)^\e\,  \G(\e -2) &= 2\, (4 \p)^{\e}\, \frac{\G (\e)}{(2 -\e)(1-\e)} \to \sdfrac{1}{\bar \e}+ \sdfrac{3}{2}+ \cO(\e)\nn
{\rm where} \qquad\sdfrac{1}{\bar \e} &\equiv \sdfrac{1}{\e} - \g_E + \ln (4 \p) = \frac{2}{4-n} - \g_E + \ln (4 \p)
\label{epsbar}
\end{align}
in which $\g_E = 0.57721...$ is the Euler-Mascheroni constant, and
\vspace{-2mm}
\begin{align}
\left(\frac{M^2}{\m^2}\right)^{\!-\e} &=  1 - \e \ln\left(\frac{M^2}{\m^2}\right) + \cO(\e^2)\nn
\left(\frac{m^2}{\m^2}\right)^{\!-\e} &=1 - \e \ln\left(\frac{m^2}{\m^2}\right)+ \cO(\e^2)\,.
\end{align}

\vspace{-6mm}
\noindent
Therefore
\vspace{-3mm}
\bes
\begin{align}
A& \to -\sdfrac{1}{\bar \e}\,  I_A  + \int_0^1\! dx\,  \ln\left(\frac{M^2}{\m^2}\right)\,  \Big[x^2 (1-x)^2 - 2\x x(1-x) + \x^2 \Big] + \cO(\e) \\[3pt]
B & \to -\left(\sdfrac{1}{\bar \e} +\sdfrac{3}{2}\right) \,  I_B + \int_0^1\! dx\,  \ln\left(\frac{M^2}{\m^2}\right)  \big(M^2)^2 
- \Big[2\x k^2m^2 +m^4\Big] \ln\left(\frac{m^2}{\m^2}\right)- \x k^2m^2 + \cO(\e) 
\end{align}
\label{ABnto4}
\ees

\vspace{-1cm}
\noindent
where the residues of the poles $I_{A,B}$ are given by the elementary integrals
\vspace{-3mm}
\bes
\begin{align}
I_A & = \! \int_0^1\! dx\, \Big[x^2 (1-x)^2 - 2\x x(1-x) + \x^2 \Big] = \sdfrac{1}{30} - \sdfrac{1}{3}\,  \x + \x^2 = \left(\x - \sdfrac{1}{6}\right)^2 + \sdfrac{1}{180}
\label{IA} \\
I_B &= \! \int_0^1\! dx\, \Big[\big(M^2\big)^2 - 2\x k^2m^2- m^4\Big] = \int_0^1\! dx\, \Big[k^4 x^2(1-x)^2 + 2k^2m^2x(1-x)- 2\x k^2m^2 \Big]\nn
& \hspace{1cm} = \sdfrac{1}{30}\, k^4 -2k^2m^2\left(\x - \sdfrac{1}{6}\right) \,. \label{IB}
\end{align}
\label{IAB}
\ees

\vspace{-1cm}
\noindent
From (\ref{STAB}), (\ref{ABnto4}) and (\ref{IAB}), the pole term in the spin-2 tensor form factor $\T(k^2)$ is given by
\bet
\T(k^2)\big\vert_{\rm pole} =-\frac{I_B}{32\p^2\, \bar\e} = 
-\frac{1}{16\p^2\, \bar \e} \left\{ \sdfrac{1}{60}\, k^4-  k^2m^2 \left(\x -\sdfrac{1}{6}\right)\right\}
\label{Tpole}
\eet
while the pole term in the spin-0 form factor $\S(k^2)$ is given by
\beo
\S(k^2)\big\vert_{\rm pole} = -\frac{\big(6 k^4 I_A - I_B\big)}{16\p^2\, \bar \e} = - \frac{1}{16\p^2\, \bar\e} \, 
\bigg\{ 6 k^4  \left(\x -\sdfrac{1}{6}\right)^2 + 2 k^2m^2  \left(\x -\sdfrac{1}{6}\right)\bigg\}\,.
\label{Spole}
\eeo
Hence both the spin-2 $\T(k^2)$ and spin-0 $\S(k^2)$ gravitational vacuum polarizations exhibit poles at $n\!=\!4$, and require UV renormalization for general 
$\x\!\neq\!1/6$. However the residue of pole term for $\S(k^2)$ of (\ref{Spole}) vanishes for the conformal value $\x\!=\!1/6$, indicating that $\S(k^2)$ is finite for
$\x\!=\!1/6$, and the $n\!\to \!4$ limit may be taken for $\S(k^2)$ directly from (\ref{STAB})-(\ref{ABdef}) in this case. This is in contrast to 
(\ref{Tpole}) for $\T(k^2)$, which is non-vanishing and requires UV renormalization for any value of $\x$.

The pole terms are removed by renormalization of local terms up to dimension $4$ in the $n\!=\!4$ gravitational action, namely
\be
{\rm S}_{\rm loc}[g] = \frac{1}{16\p G}\int \! d^4x \sqrt{-g}\, \Big(R-2\La\Big)  - \frac{1}{2} \int\!d^4x \sqrt{-g} \, 
\left(\a\, C_{abcd}C^{abcd} + \b\, R^2 \right) 
\label{Sloc}
\ee
where $C_{abcd}$ is the Weyl tensor. This local gravitational action is to be added to the effective action $\cS_{\rm 1PI}[g]$ of the quantum 
scalar $\f$ field, in order to obtain the total (1PI) gravitational effective action
\be
\cS_{\rm tot}[g] = {\rm S}_{\rm loc}[g]  + \cS_{\rm 1PI}[g]
\label{S1PI}
\ee
in which the contributions of the quantum scalar $\f$ field at all scales has been integrated out. 
The first variation of (\ref{S1PI}) produces
\be
 \frac{2\!}{\!\!\sqrt{-g}} \,\frac{\d\, \cS_{\rm tot}[g]}{\d g_{ab}(x)} = - \frac{1}{8\p G}\left (G^{ab} + \La g^{ab}\right) 
+ \a\ ^{(C)}\!H^{ab} + \b \ ^{(1)}\!H^{ab} + \big\lag \hat T^{ab}\big\rag_{\! g}
\label{S1PIvar1}
\ee
where $G^{ab}$ is the Einstein tensor and 
\vspace{-3mm}
\bes
\begin{align}
^{(C)}\!H^{ab} &= 4\,\na\mqs_c \na\mqs_d\, C^{acbd}+ 2\,R_{cd}\,C^{acbd} \\
^{(1)}\!H_{ab}&=  2\, g^{ab}\sq R  -2\na^a\na^b R + 2\, R\, R^{ab} - \sdfrac{1}{2}\, g^{ab}R^2
\end{align}\label{H1HC}
\ees
\vspace{-1cm}

\noindent
are the covariantly conserved tensors obtained from the analogous variations of the $C^2 =C_{abcd}C^{abcd}$ and $R^2$ terms in (\ref{Sloc}).
Evaluated in the flat space limit (\ref{S1PIvar1}) is 
\be
 - \frac{\La}{8\p G} \,\h^{ab} + K\,\h^{ab} =\h^{ab}\, \left[ - \frac{\La}{8\p G} 
+  (4\p)^{\!-\medmath{\sdfrac{n}{2}}}\, \G\left(\! - \sdfrac{n}{2}\right)\, \frac{m^n}{\!2}\right]
\label{Lamflat}
\ee
where (\ref{Kdef}) and (\ref{Kval}) have been used. Thus the pole term and finite part of the constant
\be
K = \frac{m^4}{32 \p^2} (4\p)^\e\,  \G(\e -2) \, \left(\frac{m^2}{\m^2}\right)^{\!-\e} = \frac{m^4}{64 \p^2} \left[ \frac{1}{\bar \e}+ \frac{3}{2}
- \ln\left(\frac{m^2}{\m^2}\right) + \cO(\e)\right]
\label{Keps}
\ee
may be absorbed into the definition of the renormalized local term 
\be
\frac{\La_R}{8\p G_R} =  \frac{\La}{8\p G} - \frac{m^4}{64 \p^2} \left[ \frac{1}{\bar \e}- \ln\left(\frac{m^2}{\m^2}\right) + \frac{3}{2}\right]
\label{LaR}
\ee
in accordance with the usual renormalization procedure~\cite{BirDav,MazurEM:1986,GorShap:2003,BuchOdinShap}.

The second variation of the total 1PI effective action (\ref{S1PI}), evaluated in flat space, is
\begin{align}
4 \ \frac{\d\, \cS_{\rm tot}[g]}{\d g_{ab}(x) \d g_{cd}(y)}\bigg\vert_{g=\h} &= - \frac{2}{8\p G}\frac {\d G^{ab}(x)}{\d g_{cd}(y)}  
- \frac{\La}{8\p G}\left(\h^{ab}\h^{cd} - \h^{ac}\h^{bd} -  \h^{ad}\h^{bc}\right)\d^4(x-y)\nn[6pt]
&+ 2 \, \left\{\a\, \frac{\d\, ^{(C)}\!H^{ab}(x)}{\d g_{cd}(y)}  + \b\,  \frac{\d\, ^{(1)}\!H^{ab}(x)}{\d g_{cd}(y)}\right\}
+ \cS_2^{abcd}(x,y)\,.
\label{Stot2var}
\end{align}
Since the metric variations by $\d g_{cd}(y)$ of the local tensors $G^{ab}(x), ^{(C)}\!H^{ab}(x), ^{(1)}\!H^{ab}(x)$ are local
derivatives of Dirac $\d$-functions, their Fourier transforms are polynomials in $k$, and in fact given by~\cite{AndMolEM:2003}
\vspace{-7mm}
\bes
\begin{align}
\int d^4\!x\ e^{ik\cdot (x-y)}\ \frac{\d\, G^{ab}(x) }{\d g_{cd}(y)} &= \sdfrac{1}{2} \,  \bigg\{ k^2 P^{(T)\,abcd}(k)  - 2 k^2 P^{(S)\,abcd}(k) \bigg\}
\label{Einvar}\\[3pt]
\int d^4\!x\ e^{ik\cdot (x-y)}\ \frac{\d\, ^{(C)}\!H^{ab}(x) }{\d g_{cd}(y)} & = -k^4  P^{(T)\,abcd}(k)\label{varHC}\\[3pt]
\int d^4\!x\ e^{ik\cdot (x-y)}\ \frac{\d\, ^{(1)}\!H^{ab}(x) }{\d g_{cd}(y)} & = -6\, k^4  P^{(S)\,abcd}(k)
\end{align}
\ees
in terms of the projection tensors of (\ref{PST}). Using the relation (\ref{Fdef}) between $F^{abcd}$ and $\cS_2$, and the result (\ref{FkP}) then gives
for the Fourier transform of (\ref{Stot2var})
\begin{align}
&4\! \int\! d^4\!x\ e^{ik\cdot (x-y)}\,\frac{\d\, \cS_{\rm tot}[g]}{\d g_{ab}(x) \d g_{cd}(y)}\bigg\vert_{g=\h} \!\!=
-\frac{k^2}{8\p G} \Big\{ P^{(T)\,abcd}(k)  - 2\, P^{(S)\,abcd}(k) \Big\} 
+\left(- \frac{\La}{8\p G} + K\right)\left(\t^{(1)\,abcd}\! - \t^{(2)\,abcd} \right)\nn[3pt]
&\hspace{2.2cm}- 2 \, k^4 \left\{\a\, P^{(T)\,abcd}(k) + 6\, \b\, P^{(S)\,abcd}(k)\right\}
- \S(k^2)\,   P^{(S)\,abcd}(k) -\T(k^2)\, P^{(T)\,abcd}(k) 
\label{Stot2k}
\end{align}

\vspace{-3mm}
\noindent
The tadpole term $K$ appears in this expression in the same combination with $-\La/8\p G$ as in (\ref{Lamflat})-(\ref{LaR}) in the first variation
of $\cS_{\rm tot}$, and can be replaced by the renormalized $-\La_R/8\p G_R$ of (\ref{LaR}).

Using (\ref{Tpole}) and (\ref{Spole}), one next observes that the pole terms in (\ref{FkP}) and (\ref{Stot2k}) proportional to one power of $k^2$
occur in the same linear combination of projectors as in the classical Einstein tensor variation (\ref{Einvar}), and therefore the terms in
\vspace{-2mm}
\begin{align}
&-\frac{k^2}{8\p G} \Big\{ P^{(T)\,abcd}(k)  - 2\, P^{(S)\,abcd}(k) \Big\}  - \frac{ \ k^2m^2}{16 \p^2 \,\bar\e} \left(\x - \frac{1}{6}\right) 
\Big\{ P^{(T)\,abcd}(k)  - 2\, P^{(S)\,abcd}(k) \Big\}\nn
&\hspace{2cm}= -\frac{k^2}{8\p \overline{G}} \Big\{ P^{(T)\,abcd}(k)  - 2\, P^{(S)\,abcd}(k) \Big\}
\label{Gren}
\end{align}
may be combined to absorb the $k^2$ pole terms into the definition of the renormalized gravitational constant
\be
\frac{1}{8\p \overline{G}} =  \frac{1}{8\p G} + \frac{m^2}{16 \p^2\, \bar\e} \left(\x - \frac{1}{6}\right) 
\label{Gbar}
\ee
in the $\overline{MS}$ scheme.

This leaves in (\ref{Stot2k}) only the pole terms proportional to $k^4$, multiplying the tensor projectors $P^{(T,S)\, abcd}(k)$. Referring to
(\ref{Tpole}), the $k^4$ pole term multiplying the spin-2 projector $P^{(T)\, abcd}(k)$ can be combined with the $\a$ term in (\ref{Stot2k}) to define
\be
\overline{\a} = \a - \frac{1}{16 \p^2\, \bar\e} \frac{1}{120}
\label{abar}
\ee
while referring to (\ref{Spole}), the $k^4$ pole term multiplying the spin-0 projector $P^{(S)\, abcd}(k)$ can be combined with the $\b$ term in (\ref{Stot2k})
to define
\beo
\overline{\b} = \b- \frac{1}{32 \p^2\, \bar\e} \left(\x - \frac{1}{6}\right)^2
\label{bbar}
\eeo
as the renormalized coefficients of the fourth order terms in (\ref{Sloc}), in the modified minimal subtraction ($\overline{MS}$) scheme. 
That the pole terms, plus finite constant terms in $1/\bar \e$ of (\ref{epsbar}) are purely local, and can be combined with local terms (\ref{Sloc}) 
in the gravitational action, shows that (\ref{LaR}), (\ref{Gbar}), (\ref{abar})-(\ref{bbar}) is a local renormalization procedure of strictly
short distance UV divergences of the 1PI effective action.

It is worth noting that the renormalization (\ref{LaR}) of the constant $\La$ term in the second variation of the 1PI effective action (\ref{Stot2k})
is simply that inherited from the first variation (\ref{Lamflat})-(\ref{LaR}) and tadpole contribution of Fig.~\ref{Fig:Tad}, with no additional subtraction 
required. Indeed no such additional subtraction is allowed, since the conservation Ward Identity (\ref{WIk}) requires the general expression (\ref{FkP}) 
to be expressed in terms of just the two tensors $P^{(T,S)\,abcd}(k)$, and there is no combination of these two transverse projection tensors proportional 
to a $k$-independent constant tensor $\t^{(1)\,abcd}\! - \t^{(2)\,abcd}$, which as a standalone tensor violates (\ref{WIk}). The impossibility of any
constant $\La$ renormalization of the second variation of the effective action (\ref{Stot2k}), beyond that required by the tadpole of Fig.~\ref{Fig:Tad}, may thus
be understood as a consequence of general coordinate invariance that gives rise to the WI of covariant conservation of $\hat T^{ab}$.

\section{Massive Field Decoupling and Physical Form Factors}
\label{Sec:Phys}

The modified minimal subtraction ($\overline{MS}$) scheme is defined by subtracting the full coefficient of $1/\bar\e$, including the finite 
$-\g_E + \ln (4 \p)$ terms in (\ref{epsbar}) in addition to the simple $1/\e$ pole in (\ref{Tpole}) and (\ref{Spole}), and then taking the
$n \to 4, \e \to 0$ limit. Hence from (\ref{ABnto4}) there remains
\vspace{-3mm}
\bes
\begin{align}
\overline A & = \int_0^1\! dx\,  \ln\left(\frac{M^2}{\m^2}\right)  \Big[x^2 (1-x)^2 - 2\x x(1-x) + \x^2 \Big] \\[3pt]
\overline B & = - \sdfrac{3}{2}\,  I_B + \int_0^1\! dx\,  \ln\left(\frac{M^2}{\m^2}\right)  \big(M^2)^2 
- \Big[2\x k^2m^2 +m^4\Big] \ln\left(\frac{m^2}{\m^2}\right)- \x k^2m^2 
\end{align}
\label{ABMSbar}
\ees

\vspace{-1cm}
\noindent
for the finite parts of the $A$ and $B$ functions. The corresponding spin-0 and spin-2 form factors $\overline \S$ and $\overline\T$ renormalized in 
the $\overline{MS}$ scheme are given by (\ref{STAB}) with $\overline A$ and $\overline B$ replacing $A$ and $B$ respectively.

As is well-known in the context of gauge theories, in general the $MS$ or $\overline{MS}$ schemes are not physical, in that quantities renormalized in such 
schemes by mass independent subtractions do not generally satisfy decoupling. This is the physical requirement, necessary for an EFT
limit, that quantum loop corrections involving fields with heavy masses should vanish when their masses $m\!\to\!\infty$~\cite{AppCar:1975}.
It is apparent by the presence of both the $\ln (M^2/\m^2)$ and $k^2m^2$ terms in (\ref{ABMSbar}) that neither function satisfies decoupling for arbitrary 
$\x$ and $\m$. However, this defect is easily remedied by identifying the terms in (\ref{ABMSbar}) proportional to $k^2m^2$ and $k^4$ that 
do not vanish as $m^2\! \to\! \infty$, and treating them as additional finite shifts in the local Newtonian $G,C^2, R^2$ couplings 
(\ref{Gbar})-(\ref{bbar}), so that the remaining terms in the spin-0 and spin-2 factors do have the required decoupling behavior 
of vanishing as $m^2 \!\to\! \infty$. As a result, any dependence of physical quantities on the arbitrary mass scale $\m$ also is removed.

In order to express the effective action in terms of physical couplings and form factors satisfying decoupling, one first writes $\ln(M^2/\m^2) 
= \ln (M^2/m^2) + \ln(m^2/\m^2)$ and notes that
\be
\ln\left(\frac{M^2}{m^2}\right) = \ln\bigg[1 + \sdfrac{k^2}{m^2} x(1-x)\bigg] = \sdfrac{k^2}{m^2} x(1-x) - \sdfrac{k^4}{2m^4} x^2(1-x)^2 
+ \cO \left(\sdfrac{k^6}{m^6}\right)
\label{log}
\ee
in terms of the physical scalar mass $m$. Then this expansion allows one to identify the terms of order $k^2m^2$ and $k^4$ in $\overline \S$ and $\overline \T$
that do not vanish as $m^2\!\to\!\infty$ and treat them as additional finite renormalizations of (\ref{Gbar})-(\ref{bbar}) instead.

Since $A$ is multiplied by $k^4$ in (\ref{STAB}), for $\overline A$ only the first step is needed and it is sufficient to write
\be
\overline A  = \int_0^1\! dx\,  \ln\left(\frac{M^2}{m^2}\right)  \Big[x^2 (1-x)^2 - 2\x x(1-x) + \x^2 \Big] + I_A \ln\left(\frac{m^2}{\m^2}\right)
\equiv A_R(k^2) + I_A \ln\left(\frac{m^2}{\m^2}\right)
\label{ARdef}
\ee
to define $A_R(k^2)$, with $I_A$ given by (\ref{IA}). On the other hand for $\overline B$ we write
\beo
\overline B = \int_0^1\! dx\,  \ln\left(\frac{M^2}{m^2}\right)  \big(M^2)^2 + I_B\, \bigg[\ln\left(\frac{m^2}{\m^2}\right) - \sdfrac{3}{2}\bigg]  -\x k^2m^2
= B_R(k^2)+ I_B \ln\left(\frac{m^2}{\m^2}\right) + 2k^2m^2 \left(\x -\sdfrac{1}{6}\right)
\label{BBR}
\eeo
with $I_B$ given by (\ref{IB}), and 
\be
B_R(k^2) \equiv \int_0^1\! dx\,  \ln\left(\frac{M^2}{m^2}\right)  \big(M^2)^2  - \sdfrac{1}{6}\, k^2m^2 - \sdfrac{1}{20}\,k^4
\label{BRdef}
\ee
defined so that the leading order $k^2m^2$ and $k^4$ terms generated by the expansion (\ref{log}) are removed from $\overline B$,
and $B_R(k^2) = \cO(k^6/m^2) \to 0$ for $m^2\!\to\!\infty$. Making use of (\ref{IAB}), and (\ref{ARdef})-(\ref{BRdef}), we now obtain
\vspace{-3mm}
\bes
\begin{align}
\overline \S &= \frac{6k^4\!}{(4 \p)^2}\, \overline A -\frac{1}{(4\p)^2}\, \overline B = \S_R(k^2) 
+ \frac{2k^2}{(4 \p)^2} \left(\x -\sdfrac{1}{6}\right)  \left\{ \bigg[3k^2\left(\x -\sdfrac{1}{6}\right)  +  m^2\bigg] \,\ln\left(\frac{m^2}{\m^2}\right) 
-  m^2 \right\}\\[3pt]
\overline \T &= \frac{1}{(4\p)^2}\, \frac{\overline B}{2} = \T_R(k^2) + \frac{k^2}{(4 \p)^2} 
\left\{ \bigg[\sdfrac{1}{60}\, k^2  - m^2 \left(\x -\sdfrac{1}{6}\right)  \bigg]\, \ln\left(\frac{m^2}{\m^2}\right)
+ m^2 \left(\x -\sdfrac{1}{6}\right) \right\}
\end{align}
\label{STbR}
\ees

\vspace{-1.1cm}
\noindent
with
\vspace{-8mm}
\bes
\begin{align}
&\S_R (k^2)= \frac{6k^4\!}{(4 \p)^2}\, A_R(k^2) -\frac{1}{(4\p)^2}\,B_R(k^2)\nn[5pt] 
&\hspace{-1cm}=\frac{1}{16\p^2}\!\int_0^1\!dx\, \ln\left(\frac{M^2}{m^2}\right) \bigg\{k^4\, \Big[5x^2(1-x)^2 - 12 \x x(1-x) + 6\x^2\Big]
-2k^2m^2 x(1-x) - m^4\bigg\}\nn
&\hspace{3cm}+\frac{1}{16\p^2} \left(\sdfrac{1}{20}\,  k^4 + \sdfrac{1}{6} k^2m^2 \right)\label{SR}\\[6pt]
&\T_R(k^2) = \frac{1}{(4\p)^2}\, \frac{B_R(k^2)}{2} 
=\frac{1}{32 \p^2} \left\{\int_0^1\!dx\,\ln\left(\frac{M^2}{m^2}\right) \big(M^2)^2 -\sdfrac{1}{20}\,  k^4 - \sdfrac{1}{6} k^2m^2\right\}
\label{tauR}
\end{align}
\label{SRTR}
\ees
both of which vanish as $k^6/m^2$ for $m^2\!\to\infty$.

The additional finite terms in (\ref{STbR}) contribute 
\vspace{-3mm}
\begin{align}
&\frac{1}{(4 \p)^2}\, k^2m^2 \left(\x -\sdfrac{1}{6}\right)  \bigg[\ln\left(\frac{m^2}{\m^2}\right) - 1\bigg]
 \Big\{ P^{(T)\,abcd}(k)  - 2\, P^{(S)\,abcd}(k) \Big\}\nn[3pt]
&   -\frac{1}{(4 \p)^2} \frac{\,k^4\!}{60}\,\ln\left(\frac{m^2}{\m^2}\right) \,  P^{(T)\,abcd}(k)  
-  \frac{6 k^4}{(4 \p)^2}\left(\x -\sdfrac{1}{6}\right)^2\, P^{(S)\,abcd}(k)
\end{align} 
to (\ref{Stot2k}). As in (\ref{Gren})-(\ref{bbar}) these terms can be combined with the second variations of local terms from the 
gravitational action (\ref{Sloc}) that have the same tensor structure, to define the fully renormalized physical parameters
\bes
\begin{align}
\frac{1}{8\p G_R} &= \frac{1}{8 \p \bar G} - \frac{m^2}{(4 \p)^2} \left(\x -\sdfrac{1}{6}\right)  \bigg[\ln\left(\frac{m^2}{\m^2}\right) - 1\bigg]
=  \frac{1}{8 \p G} + \frac{m^2}{16\p^2} \left(\x -\sdfrac{1}{6}\right)  \left[ \frac{1}{\bar\e}- \ln\left(\frac{m^2}{\m^2}\right) + 1 \right]\\[6pt]
&\a_R(0)  = \overline\a + \frac{1}{(4 \p)^2} \frac{1}{120} \,\ln\left(\frac{m^2}{\m^2}\right)  
= \a -\frac{1}{1920 \p^2} \left[\frac{1}{\bar \e} - \ln\left(\frac{m^2}{\m^2}\right) \right]\\[6pt]
&\b_R(0) = \overline\b + \frac{1}{2(4\p)^2} \left(\x -\sdfrac{1}{6}\right)^2  = \b 
- \frac{1}{32\p^2} \left(\x -\sdfrac{1}{6}\right)^2 \left(\sdfrac{1}{\bar\e} -1\right)
\end{align}
\label{GabR}
\ees

\vspace{-1cm}
\noindent
at $k^2\!=\! 0$. The result is that the full second variation of the 1PI effective action of (\ref{Stot2k}) can be expressed 
\begin{align}
&4\! \int\! d^4\!x\ e^{ik\cdot (x-y)}\,\frac{\d\, \cS_{\rm tot}[g]}{\d g_{ab}(x) \d g_{cd}(y)}\bigg\vert_{g=\h} \!\!=
- \frac{\La_R}{8\p G_R}\left(\t^{(1)\,abcd}\! - \t^{(2)\,abcd} \right) -\frac{k^2}{8\p G_R} \bigg\{ P^{(T)\,abcd}(k)  - 2\, P^{(S)\,abcd}(k) \bigg\} \nn[6pt]
&\hspace{5mm}- 2 \, k^4\, \bigg\{\a_R(0) \, P^{(T)\,abcd}(k) + 6\, \b_R(0) \, P^{(S)\,abcd}(k)\bigg\}
- \S_R(k^2)\,   P^{(S)\,abcd}(k) -\T_R(k^2)\, P^{(T)\,abcd}(k) 
\label{Stot2R}
\end{align}
entirely in terms of physical renormalized parameters (\ref{GabR}) at $k^2\!=\!0$, and physical renormalized form factors (\ref{SRTR})
that are functions of $k^2$ vanishing at $k^2\!=\!0$ and satisfying decoupling. Both $\S_R(k^2)$ and $\T_R(k^2)$ are also independent of
the arbitrary mass scale $\m$ introduced in dimensional regularization.

The renormalized form factors (\ref{SRTR}) may be used to define the physical dimensionless response functions $F^{(S,T)}$ of $k^2/m^2$ by
\vspace{-5mm}
\bes
\begin{align}
\S_R (k^2)& = k^4 F^{(S)} \left(\sdfrac{k^2}{m^2}\right)\\[3pt]
\T_R(k^2) &= k^4 F^{(T)} \left(\sdfrac{k^2}{m^2}\right)
\end{align}
\label{Fphys}
\ees

\vspace{-1.2cm}
\noindent
with
\bet
F^{abcd}_R(k) = -\Big[ k^4  P^{(S)\,abcd}(k)\Big]\, F^{(S)} \left(\sdfrac{k^2}{m^2}\right) -\Big[k^4 P^{(T)\,abcd}(k)\Big] \, F^{(T)} \left(\sdfrac{k^2}{m^2}\right)
\label{FR}
\eeo
the renormalized polarization tensor, after the local contact terms in (\ref{Stot2R}) have been removed. The finite integrals for the two response 
functions defined by (\ref{SRTR}) and (\ref{Fphys}) can be evaluated in closed form by
\vspace{-3mm}
\begin{align}
F^{(S)} \left(\sdfrac{k^2}{m^2}\right) &= \frac{1}{16\p^2}\!\int_0^1\!dx\, \ln\left(\sdfrac{M^2}{m^2}\right) \bigg[5x^2(1-x)^2 - 12 \x x(1-x) + 6\x^2
- \sdfrac{2m^2}{k^2} x(1-x) - \sdfrac{m^4}{k^4}\bigg]\nn
&\hspace{3cm}+\sdfrac{1}{16\p^2} \left(\sdfrac{1}{20} + \sdfrac{m^2}{6\,k^2} \right)\nn[6pt]
&\hspace{-1.2cm} =  \sdfrac{1}{144\p^2}\bigg\{\sdfrac{1}{10}+ \big(1-6\x\big)   -  \sdfrac{m^2}{k^2}
+ 3\,\bigg[\big(1-6\x\big)- \sdfrac{2m^2}{k^2}\bigg]^2\ \bigg[z\, {\rm tanh}^{-1}  \Big(\sdfrac{1}{z}\Big) -1\bigg]\bigg\}
\label{FS}
\end{align}
which coincides with eq.~(B45) of Ref.~\cite{AndMolEM:2003}, when it is recognized that
\be
2z\, {\rm tanh}^{-1}\, \Big(\sdfrac{1}{z}\Big) = z\, \ln \left(\sdfrac{z+1}{z-1}\right) = f\left(\sdfrac{k^2}{m^2}\right) \quad {\rm for} \quad 
z= \sqrt{1 + \sdfrac{4m^2}{k^2}} > 1
\label{ffn}
\ee
in the notation of \cite{AndMolEM:2003} for this function $f$. Likewise
\begin{align}
F^{(T)} \left(\sdfrac{k^2}{m^2}\right) &
= \frac{1}{32 \p^2} \left\{\int_0^1\!dx  \ln\left(\sdfrac{M^2}{m^2}\right)\left[x^2(1-x)^2 + \sdfrac{2m^2}{k^2}x(1-x) + \sdfrac{m^4}{k^4}\right] -\sdfrac{1}{20}- \sdfrac{m^2}{6k^2}\right\}\nn[3pt]
&\hspace{-1cm}
= \frac{1}{480\p^2} \left\{z^5 {\rm tanh}^{-1}  \Big(\sdfrac{1}{z}\Big) -  z^4 - \sdfrac{1}{3}\, z^2  -\sdfrac{1}{5}\right\}
\quad {\rm for} \quad  z= \sqrt{1 + \sdfrac{4m^2}{k^2}} > 1\,.
\label{FT}
\end{align}
which agrees with eq.~(B44) of Ref.~\cite{AndMolEM:2003}, with the use of (\ref{ffn}).

The physical response functions (\ref{Fphys}) also allow the definition of renormalized couplings $\a_R$ and $\b_R$ as functions of 
physical external momentum $k^2 >0 $, by
\vspace{-3mm}
\bes
\begin{align}
\a_R(k^2) &= \a_R(0) + \sdfrac{1}{2}\, F^{(T)}\left(\sdfrac{k^2}{m^2}\right)\\[4pt]
\b_R(k^2) &= \b_R(0) + \sdfrac{1}{12} \, F^{(S)}\left(\sdfrac{k^2}{m^2}\right)
\end{align}
\label{aRbR}\ees

\vspace{-1cm}
\noindent
with $\a_R(0), \b_R(0)$ their values at $k^2 =0$. Their logarithmic running with the momentum scale $k^2$ 
\vspace{-2mm}
\begin{align}
k^2 \sdfrac{d}{dk^2}\, \a_R(k^2) &= \sdfrac{1}{2}\,k^2 \sdfrac{d}{dk^2} F^{(T)}\left(\sdfrac{k^2}{m^2}\right)\label{arun}\\[6pt]
k^2 \sdfrac{d}{dk^2} \,\b_R(k^2) &= \sdfrac{1}{12}\, k^2 \sdfrac{d}{dk^2}F^{(S)}\left(\sdfrac{k^2}{m^2}\right)\label{brun}\,.
\end{align}
is independent of the arbitrary scale $\m$ introduced in the MS or $\overline{MS}$ scheme. This eliminates any need
to supply that arbitrary mass scale $\m$ with a physical interpretation. Because $F^{(S,T)}$ satisfy decoupling, vanishing 
as $k^2/m^2$ for $k^2/m^2 \to 0$, the running of the renormalized couplings in (\ref{arun})-(\ref{brun}) also go to zero in this low 
momentum or heavy mass limit, from (\ref{FSTdec}), and are in agreement with those given in Ref.~\cite{GorShap:2003}. 
In contrast the gravitational constant $G$ and cosmological term $\La$ do not acquire any running with the external momentum
scale $k^2$, neither when renormalized in the $\overline{MS}$ scheme, as noted in \cite{GorShap:2003,ShapSol:2009},
nor by the additional finite subtractions in (\ref{GabR})-(\ref{Stot2R}) to enforce decoupling.

\section{The Conformal Anomaly in the Massless Limit}
\label{Sec:Anom}

As already noted in Sec.~\ref{Sec:MS}, from (\ref{Spole}), the residue of the spin-0 polarization function at $n\!=\!4$ vanishes
identically for $\x\! = \!1/6$, as do the additional finite subtractions in passing from $\overline \S$ to the physical renormalized
$\S_R(k^2)$ in (\ref{SR}). Hence $\S = \bar \S = \S_R$ is finite, there is no UV renormalization for $\S$ to be performed, and this finite $\S$ 
also satisfies the physical decoupling requirement automatically in the $\x\!=\!1/6$ case. This allows its behavior in the infrared EFT limit 
be investigated directly and unambiguously, without any dependence upon regularization, renormalization or UV physics.

Setting $\x \!=\!1/6$ in (\ref{FS}) gives the relatively simple compact result
\bet
F^{(S)}\left(\sdfrac{k^2}{m^2}\right)\bigg\vert_{\x=1/6}= \frac{1}{16\p^2} \,\frac{1}{90}\,  \left\{1 - \frac{10\, m^2}{\,k^2} 
+ \frac{60\,m^4}{\!k^4}\,\left[ z\, \ln \left(\frac{z+1}{z-1}\right) -2 \right]\right\}
\label{FSconf}
\ee
for $z= \!\sqrt{1 + \sdfrac{4m^2}{k^2}},\ k^2 > 0$. This spin-0 response function has the limits
\be
F^{(S)}\left(\sdfrac{k^2}{m^2}\right)\bigg\vert_{\x=1/6}\to \frac{1}{16\p^2}\frac{1}{90}\times 
\left\{\begin{array}{ccc}{\displaystyle \sdfrac{1}{7^{\vphantom{(}}} \sdfrac{k^2}{m^2}}&\to 0
\quad& m^2\to \infty\\[4pt]
1 & &m^2 \to 0\end{array}\right.
\label{FSlims}
\ee
satisfying decoupling directly in the $m^2 \to \infty$ limit, with no subtraction or amendment. 

In the opposite massless limit, the finite value of (\ref{FSlims}) at $m^2\!=\!0$ determines the coefficient of the $\sq R$ term in the conformal trace anomaly 
of the renormalized $\lag \hat T^{ab}\rag_R$. For a massless, conformally coupled scalar field the conformal anomaly is~\cite{Brown:1977,Duff:1977,BirDav}
\vspace{-3mm}
\beo
\big\lag \hat T^a_{\ a}\big\rag_R\,\Big\vert_{\x=1/6\,,m=0}= \frac{1}{16\p^2}\left\{- \frac{1}{360}\,E + \frac{1}{120} C_{abcd}C^{abcd} +  \frac{1}{180}\sq R\right\}
\label{anomphi}
\ee
in a general four dimensional curved spacetime, with $E= R_{abcd}R^{abcd} - 4R_{ab}R^{ab} + R^2$. Since the first two terms in (\ref{anomphi}) are 
quadratic in the curvature tensor, their first variation evaluated in flat space vanishes. However the variation of the third term survives, with
\beo
\frac{\d\, \big\lag \hat T^a_{\ a}(x)\big\rag_R}{\d g_{cd}(y)} =  \frac{1}{(4\p)^2}\frac{1}{180}\sq\left( \frac{\d R(x)}{\d g_{cd}(y)}\right)
=  \frac{1}{16\p^2}\frac{1}{180}\sq \left(\na^c \na^d - \h^{cd} \sq\right)\d^4(x-y)
\label{vartr}
\eeo
when evaluated around flat space. On the other hand, from (\ref{var2S}) and (\ref{Fdef}), we have that
\beo
\frac{\d\, \big\lag \hat T^a_{\ a}(x)\big\rag_R }{\d g_{cd}(y)}  = \frac{1}{2}\, \h_{ab}\, F^{abcd}_R(x,y)\Big\vert_{\x=1/6\,,m=0}
\label{varTF}
\eeo
since the $K$ dependent terms proportional to $\big\lag \hat T^{ab}\big\rag$ in (\ref{Fdef}) cancel from this variation. Fourier transforming 
(\ref{varTF}), and using (\ref{thPS}), (\ref{FR}), and (\ref{FSlims}) gives
\be
\frac{1}{2} \h_{ab}\,F^{abcd}_R(k)\Big\vert_{\x=1/6\,,m=0}= -\frac{1}{2} \h_{ab}P^{(S)abcd}(k)\,k^4 F^{(S)}\left(\sdfrac{k^2}{m^2}\right)\bigg\vert_{\x=1/6, m=0}  
= \frac{1}{16\p^2}\frac{1}{180}\, k^2 \left(k^c k^d - k^2 \h^{cd}\right) 
\label{Siganom}
\eeo
in momentum space, which is exactly the Fourier transform of (\ref{vartr}). Thus the coefficient of (\ref{FSlims}) in the $m\!\to\!0$ limit is precisely 
twice that of the $\sq R$ term in the conformal anomaly (\ref{anomphi}), as noted also in \cite{GorShap:2003}.

\begin{figure}[h]
\vspace{-1mm}
\includegraphics[height=7.5cm,width=11cm, trim=0cm 0cm 0cm 0cm, clip]{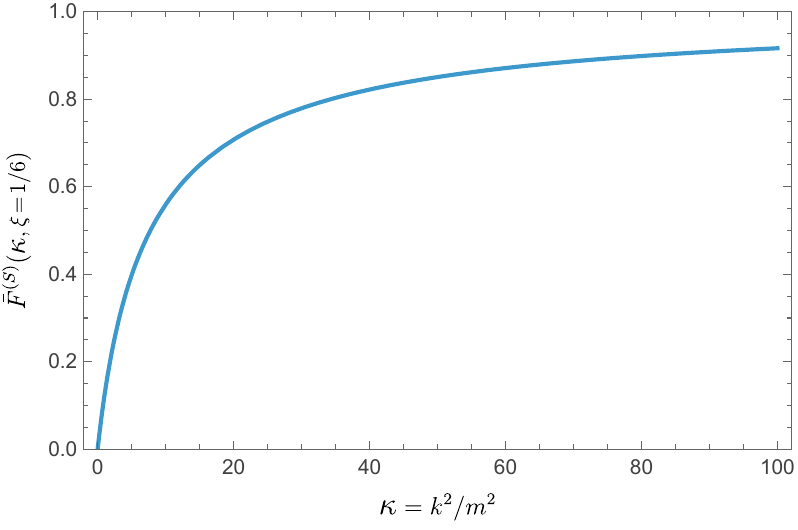}
\vspace{-2mm}
\singlespace
\caption{The rescaled spin-0 response function $\bar F^{(S)}\Big(\scalebox{1.2}{$\ka$}, \scalebox{.9}{$\x \!=\!1/6$}\Big) = (16\p^2)(180)\,  
F^{(S)}(\scalebox{1.2}{$\ka$})\vert_{\x = 1/6}$, eq.~(\ref{FSconf}), as a function of $\scalebox{1.2}{$\ka$} = k^2/m^2$, varying 
smoothly between zero at $\ka =0$ and unity as $\ka \to \infty$.}
\label{Fig:FS}
\vspace{-3mm}
\end{figure}

The rescaled spin-0 response function $(16\p^2)(180)F^{(S)}$ is shown as a function of $\ka =k^2/m^2$ in Fig.~\ref{Fig:FS}. Since the value of this 
function in the massless limit, $\ka\!\to\!\infty$, gives the $\sq R$ term in the trace anomaly, the $\sq R$ anomaly coefficient is unambiguously fixed 
by the IR behavior of gravitational vacuum polarization of the scalar field at $m=0$, independently of UV regularization or renormalization. Although at $m=0$ 
the anomaly $\sq R$ is local, the effective action that gives rise to it, namely
\vspace{-3mm}
\begin{align}
&- \sdfrac{1}{2^2 2!} \!\int\! d^4\!x\ h_{ab}(x)\int \!d^4\!y\ h_{cd}(y) \int \frac{d^4 k}{(2 \p)^4}  e^{ik\cdot (x-y)} P^{(S)\,abcd}(k) \,\S(k^2)\big\vert_{\x = 1/6}
\nn
&= -\sdfrac{1}{24} \!\int\! d^4\!x\  \d R(x)\!\int \!d^4\!y\  \d R(y)\! \int\! \frac{d^4 k}{(2 \p)^4}\,  e^{ik\cdot (x-y)}\, F^{(S)}\left(\sdfrac{k^2}{m^2}\right)\bigg\vert_{\x = 1/6}
\label{FSaction} 
\end{align}
is non-local for any value of $m \neq 0$, and not removable by any local counterterm. 

\begin{figure}[h]
\vspace{-2mm}
\includegraphics[height=7cm,width=11.5cm, trim=0cm 0cm 0cm 0cm, clip]{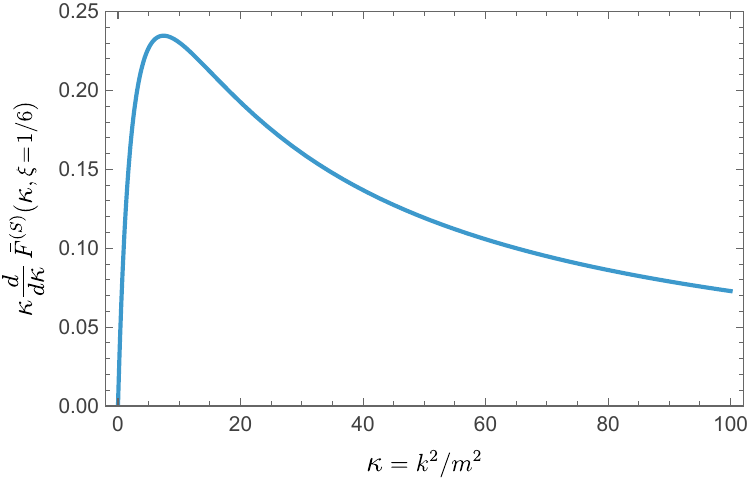}
\vspace{-3mm}
\singlespace
\caption{The logarithmic derivative of the rescaled spin-0 response function $\bar F^{(S)}\Big(\scalebox{1.2}{$\ka$}, \scalebox{.9}{$\x \!=\!1/6$}\Big)$ of Fig.~\ref{Fig:FS}
determining the running of the coupling $\b$ by eq.~(\ref{brun}).} 
\label{Fig:dFS}
\vspace{-4mm}
\end{figure}

The logarithmic derivative of $F^{(S)}$ determining the physical running coupling \,$\b_R(k^2)$ by (\ref{brun})  is non-zero for any finite 
$\ka = k^2/m^2 \in (0, \infty)$  as shown in Fig.~\ref{Fig:dFS}, but vanishes at both endpoints of this interval. The rate of this finite infrared renormalization
of the rescaled $F^{(S)}$ is shown in Fig.~\ref{Fig:dFS}.

Retaining a finite, non-zero mass $m$, or equivalently $\ka = k^2/m^2$ finite, is essential for distinguishing UV from IR physics at the two limits 
of (\ref{FSlims}). With decoupling satisfied, $m^2$ separates the momentum scales of quantum $\f$ field modes 
which are integrated out in the effective action ($k^2 \gtrsim m^2$) from those which are not integrated out ($k^2 \lesssim m^2$). The response function
(\ref{FSconf}), plotted in Fig.~(\ref{Fig:FS}) may be regarded as the variation of the non-local $\d R(x)\, \d R(y)$ term in the effective action (\ref{FSaction}) 
as a function of the {\it infrared} cutoff $L^2 =1/m^2$, with fixed UV behavior at $L=0$, as $L$ varies from $0$ to $\infty$. This realizes the essential aspect 
of Wilsonian renormalization, but in the 1PI effective action formulation of continuum QFT, with $1/m$ acting as smooth and Lorentz invariant IR distance cutoff 
scale. Taking $m \!\to\! 0$ or equivalently $\ka = k^2/m^2\! \to \!\infty$ in the second limit of (\ref{FSlims}), where the $\f$ field modes at all scales are taken 
into account, results in the standard conformal anomaly of the $\sq R$ term in the far infrared, with no dependence whatsoever on UV regularization or
renormalization effects.

\section{Imaginary Parts and Dispersive Representation}
\label{Sec:ImSpec}

For $-4m^2 < k^2 <0$, $z^2 < 0$, $s = -k^2 >0$ for $k^2 <0$ timelike, and $z= \pm i\, |z|$ becomes purely imaginary. 
However, both $F^{(S)}$ and $F^{(T)}$ remain real , with
\be
z\, {\rm tanh^{-1}}\Big(\sdfrac{1}{z}\Big) = |z|  \, {\rm tan^{-1}}\bigg(\sdfrac{1}{|z|}\bigg) \,,\qquad |z| = \sqrt{\frac{4m^2}{\!s}- 1}
\quad {\rm for}\quad s= -k^2 \quad {\rm and}\quad 0<s < 4m^2.
\ee
At $k^2\! =\! 0$, the response functions are analytic, in fact vanishing, behaving as
\bes
\begin{align}
F^{(S)} \left(\sdfrac{k^2}{m^2}\right) & =\frac{1}{144\p^2} \frac{k^2}{m^2}\, \bigg[\frac{1}{70} -\frac{1}{10} \,\big(6\x-1\big)
+\frac{1}{4}\, \big(6 \x-1\big)^2\bigg]  + \cO\left(\frac{k^4}{m^4}\right)\quad \to\quad  0\\[6pt]
F^{(T)} \left(\sdfrac{k^2}{m^2}\right) & = \frac{1}{480\p^2}\left\{ \frac{1}{7z^2} + \cO\big(1/z^4\big)\right\} =  
\frac{1}{192\p^2}\, \frac{1}{70} \,\frac{k^2}{m^2}+ \cO\left(\frac{k^4}{m^4}\right)\quad  \to \quad 0\nn[-1.2cm]
\end{align}
\label{FSTdec}
\ees
and hence satisfy decoupling as $k^2/m^2\!\to\! 0$ in the heavy mass limit.

For $s =-k^2 > 4m^2$, $0 < z< 1$, the response functions have a logarithmic branch cut and acquire an imaginary part above the threshhold for
creation of a real two-particle pair. The case of retarded response functions was treated in Ref.~\cite{AndMolEM:2003}, in which case
\be
f_{\rm ret}\left(\sdfrac{k^2}{m^2}\right)= z \ln\left(\sdfrac{1+z}{1-z}\right) - i \p \, z\,  {\rm sgn}(k^0)\, \Th (s-4m^2)  \qquad {\rm with} \qquad 
0< z= \sqrt{1 - \frac{4m^2}{s}} < 1\, .
\label{fcut}
\ee
For response functions obeying the Feynman $-i \e$ prescription for the imaginary part, the ${\rm sgn}(k^0)$ factor in  (\ref{fcut}) is absent.

The imaginary parts of the response functions $F^{(S,T)}$ can be obtained directly from their integral representations (\ref{Fig:FS}),
arising when the argument of the logarithm $M^2/m^2 = 1 + k^2 x(1-x)/m^2$ becomes negative. This occurs when 
\be
s = -k^2 \ge \frac{m^2}{x(1-x)} \ge 4m^2
\label{srange}
\ee
and for $x_- \le x \le x_+$ where 
\beo
x_{\pm} = \sdfrac{1}{2} \big( 1 \pm r\big)\,, \qquad r \equiv \sqrt{1 - \frac{4m^2}{s}}
\eeo
are the two real roots of $x_{\pm}(1-x_{\pm}) = m^2/s$. 

Choosing the logarithm branch in accordance with the $m^2-i \e$ prescription for the Feynman response function,
we obtain for the imaginary parts
\be
{\rm Im}\ F^{(S,T)}\left(\sdfrac{k^2}{m^2}\right) = - \sdfrac{\p}{s^2}\, \r^{(S,T)}(s)\qquad {\rm for}\quad  k^2 = -s\quad {\rm timelike}\,.
\label{ImST}
\ee
where $\r^{(S,T)}(s)$ are the positive semi-definite functions,
\vspace{-2mm}
\bes
\begin{align}
\r^{\rm (S)}(s)&= \frac{\Th(s-4m^2)} {24 \p^2}\sqrt{1-\frac{4m^2}{s}} \,\left[\frac{(1-6 \x)s}{2} + m^2 \right]^2\label{rS}\\[6pt]
\r^{(\rm T)}(s)&= \frac{\Th(s-4m^2)}{60 \pi^2}\sqrt{1-\frac{4m^2}{s}}\, \left(\frac{s}{4}-m^2\right)^2\label{rT}
\end{align}
\label{rST}
\ees

\vspace{-1cm}
\noindent
in agreement with Eqs.~(B37) of \cite{AndMolEM:2003}. In obtaining (\ref{ImST})-(\ref{rST})
\be
\int_{x_-}^{x_+}\!dx\, x(1-x) = \frac{r}{6}\,  \left(1 + \sdfrac{2m^2}{s}\right)\,,\qquad 
\int_{x_-}^{x_+}\!dx\, x^2(1-x)^2 = \frac{r}{30}\, \left(1 + \frac{2m^2}{s^{\phantom{(}}} + \frac{6m^4}{s^2}\right)
\ee
have been used. 

It is straightforward to show by direct integration that both response functions (\ref{FS}) and (\ref{FT}) may also be expressed as dispersion intergrals
\beo
F^{(S,T)} \left(\sdfrac{k^2}{m^2}\right) = k^2 \!\int_0^\infty \frac{ds}{s^3\big(s+k^2 - i\e\big)} \ \r ^{(S,T)}(s)
\label{FSTdis}
\vspace{2mm}
\eeo
with $\r ^{(S,T)}$ given by (\ref{rST}). This is in agreement with \cite{AndMolEM:2003}, where only the imaginary parts (\ref{rST}) were first computed, 
and the real parts of (\ref{FSTdis}) inferred by causality and the (thrice subtracted) dispersion relations (\ref{FSTdis}) were obtained.

The triply subtracted dispersion relations of \cite{AndMolEM:2003} are justified by UV renormalization of counterterms in the local action (\ref{Sloc}) 
up to dimension four, removing quartic, quadratic and logarithmic UV divergences, that are generically present for $\x\!\neq\!1/6$.  This mass dependent 
subtraction procedure guarantees that the physical requirement of decoupling of the response functions is satisfied as $m^2 \!\to\! \infty$, unlike in 
the mass independent $\overline{MS}$ subtraction scheme. Once amended by the additional finite subtractions of local terms in (\ref{GabR}) 
and (\ref{Stot2R}), the physical polarization and response functions do satisfy decoupling, and when so amended, the $\overline{MS}$ scheme then 
agrees completely with the dispersive approach of Ref.~\cite{AndMolEM:2003}. In the $\x\!=\!1/6$ case, no additional subtractions of any kind 
are required for the spin-0 polarization or response function $F^{(S)}$.

\section{Spin-0 Spectral Function and UV Finite Sum Rule for $\x = 1/6$}
\label{Sec:SpecSum}

For $\x\!=\!1/6$, from (\ref{rS}), one may represent the spin-0 spectral function appearing in (\ref{FSTdis})  
\vspace{-3mm}
\begin{align}
\s(s)& \equiv \frac{1}{s^3}\,\r^{(S)}(s) \Big\vert_{\x = 1/6} =\frac{m^4\!}{24\p^2\, s^3}\sqrt{1 - \frac{4m^2}{s}}\, \Theta\big(s-4m^2\big) \nn[6pt]
&= \frac{1}{48\p^2} \!\int_0^1 dx\, x(1-x) (1-2x)^2\,  \d\left(s - \sdfrac{m^2}{x(1-x)}\right)
\label{sigdef}
\end{align}
by the latter integral over $x$, as may be checked by its direct evaluation. By substitution of this representation into (\ref{FSTdis}), interchange of
the $s$ and $x$ integrals, and integrating by parts, one finds
\begin{align}
&F^{(S)} \left(\sdfrac{k^2}{m^2}\right)\bigg\vert_{\x = 1/6}\!\! = k^2 \!\int_0^\infty \frac{ds\ \s(s)}{ \big(s+k^2 - i\e\big)} =
\frac{k^2\!}{48\p^2} \!\int_0^1 dx\, \frac{x^2(1-x)^2 (1-2x)^2}{m^2 + k^2 x(1-x)}\nn[3pt]
&= \frac{1}{48\p^2} \!\int_0^1 dx\, x^2(1-x)^2 (1-2x)\, \frac{\pa}{\pa x} \ln \left(\sdfrac{M^2}{m^2}\right)\nn[3pt]
&\hspace{5mm}= -\frac{1}{48\p^2} \!\int_0^1 dx\, \ln \left(\sdfrac{M^2}{m^2}\right)  \frac{d}{dx}\, \Big[x^2(1-x)^2 (1-2x)\Big]\nn[3pt]
&= \frac{1}{24\p^2} \!\int_0^1 dx\,\ln \left(\sdfrac{M^2}{m^2}\right)\,  \Big[5 x^2(1-x)^2 - x(1-x)\Big]
\label{FSconfsim}
\end{align}
which once again evaluates to (\ref{FS}) for $\x\!=\!1/6$.

In addition to (\ref{FSconfsim}) providing perhaps the simplest integral representation of spin-0 response function (\ref{FS}), it follows immediately from (\ref{sigdef})
by interchanging the $s$ and $x$ integrals that the spectral function $\s(s)$ satisfies the UV finite sum rule
\beo
\int_0^\infty ds \,\s(s) =  \int_0^\infty \frac{ds}{s^3} \,\r^{(S)}(s)\big\vert_{\x = 1/6} 
= \frac{1}{48\p^2} \!\int_0^1 dx\, x(1-x) (1-2x)^2 = \frac{1}{16\p^2}\frac{1}{90} =  \frac{1}{1440\p^2}
\label{sumrule}
\eeo
which from (\ref{FSlims}) is just the value of $F^{(S)} \left(k^2/m^2\right)\big\vert_{\x = 1/6}$  at $m^2=0$, and simply related to the 
coefficient of the $\sq R$ term in the conformal anomaly by (\ref{vartr})-(\ref{Siganom}). 

\begin{figure}[h]
\includegraphics[height=7cm,width=12cm, trim=0cm 0cm 0cm 0cm, clip]{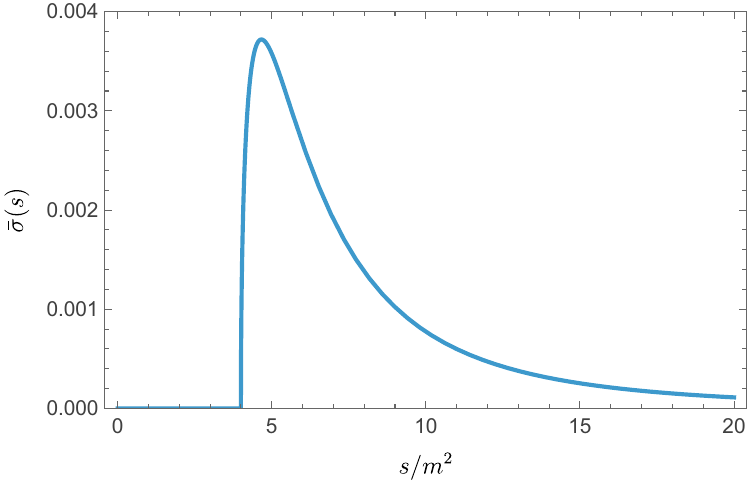}
\vspace{-1mm}
\singlespace
\caption{The rescaled spectral function $\bar \s(s) = (24 \p^2 m^2)\, \s(s)$, (\ref{sigdef}) as a function of $s/m^2$.}
\label{Fig:Spec}
\vspace{-3mm}
\end{figure}

The sum rule (\ref{sumrule}) holds for any $m$, while $\s(s)$ vanishes pointwise for all $s > 0$ in the limit $m^2\to 0$. This is compatible with 
(\ref{sumrule}) only by $\s(s)$ becoming a singular Dirac $\d$-function at $s=0$ in this massless limit, with the sum rule saturated by the $\d$-function. 
Indeed this follows immediately from the integral representation (\ref{sigdef}) for $\s(s)$, which together with (\ref{sumrule}) gives
\be
\s(s)\big\vert_{m=0} = \frac{1}{16 \p^2}\, \frac{1}{90}\, \d(s)
\label{sigdel}
\ee
for $m=0$. The function $\bar\s(s) = 24\p^2 m^2 \s(s)$ is shown in Fig.~\ref{Fig:Spec} as a function of $s/m^2$. It rises sharply from zero at threshold $s=4m^2$ 
to a maximum at $s= 14 m^2/3$ where the value of $\bar \s$ is $(3/14)^3/\!\sqrt{7} = 3.719 \times 10^{-3}$. It then falls asymptotically like $1/s^3$, 
with the full width in $s$ at half maximum of the peak being $\sim 3m^2$. Thus the peak in $\s(s)$ approaches $s\!=0\!$ as the threshhold $4m^2 \!\to\!0$
and its height grows $\propto 1/m^2$, with narrower and narrower width $\propto m^2$ as $m^2\! \to\! 0$. The result is that the area under 
the curve in Fig.~\ref{Fig:Spec} given by (\ref{sumrule}) is independent of $m$, while $\s(s)$ approaches the Dirac $\d$-function of (\ref{sigdel}),
as  $m^2\! \to\! 0$.

Both the existence of a UV finite sum rule (\ref{sumrule}) and the appearance of a Dirac $\d$-function in the spectral function in the massless limit 
are characteristic features of anomalous amplitudes~\cite{GiaEM:2009}, with the observation that the Adler axial anomaly in QED is due to exactly
such a $\d$-function at zero center-of-mass energy, originally recognized in \cite{DolZak:1971}. See also \cite{Horejsi:1985}. As in the case of 
the axial anomaly, these features are independent of any UV physics, divergences or regulators, provided only that coordinate and Lorentz invariance 
are respected by the quantum effective action. The Dirac $\d$-function in the imaginary part corresponds directly with a $1/k^2$ massless propagating 
intermediate state in the spin-0 trace projection of the cut polarization diagram, since 
\be
\int_0^\infty \frac{ds}{k^2 + s} \, \s(s) \ \stackrel{m^2\to 0}{\to}\ \frac{1}{16 \p^2}\, \frac{1}{90} \frac{1}{k^2}
\label{realsig}
\ee
with the residue of the $1/k^2$ pole given by (twice) the coefficient of the $\sq R$ term in the conformal anomaly (\ref{anomphi}). 

It is instructive to compare this with the two-dimensional case, which is provided in Appendix \ref{App:2D}. The fact that there is an additional factor of $k^2$ 
in (\ref{FSconfsim}), corresponding to $\sq R$ rather than just $R$ in the two-dimensional conformal anomaly, results in only the residue of the pole 
in (\ref{realsig}) remaining in $F^{(S)}$, while the pole itself does not appear in the 4D two-point function. The general form of the conformal 
anomaly effective action in $n=4$ shows that the conformal anomaly pole in four dimemsions appears first in the three-point function $\cS_3^{abcdef}$, 
involving three SET correlations, corresponding to the triangle diagram of the Adler axial anomaly~\cite{TTTCFT:2019}.

For comparison one may consider the spin-2 tensor response function (\ref{FT}) which may also be written
\beo
F^{(T)}\left(\sdfrac{k^2}{m^2}\right) =  \frac{1}{960\p^2} \left\{ - \frac{46}{15} - \frac{56}{3} \frac{m^2}{k^2} - \frac{32\, m^4}{k^4} 
+  z^5\,\ln\left(\frac{z+1}{z-1}\right)\right\} \label{FTa}
\eeo
coinciding with eq.~(B44) of Ref.~\cite{AndMolEM:2003}, obtained by the dispersion method, with the spin-2 spectral function (\ref{rT}).
Unlike the spin-0 function the spin-2 spectral function grows with $s^2$ as $s \to \infty$, so that the integral of $\r^{(T)}/s^3$ is logarithmically
divergent and does not satisfy a UV finite sum rule. Correspondingly, the real part (\ref{FT}) or (\ref{FTa}) behaves logarithmically
\be
F^{(T)}\left(\sdfrac{k^2}{m^2}\right) \stackrel{k^2\to\infty}{\longrightarrow} \frac{1}{960 \p^2} \, \ln \left(\sdfrac{k^2}{m^2}\right)
\label{logT}
\ee
as $k^2 \!\to\! \infty$ or $m^2\! \to\! 0$, rather than tending to a finite value as $F^{(S)}$ for $\x\!=\!1/6$ does.
There is no massless pole at $k^2=0$ for $m^2\!=\!0$ in the spin-2 traceless sector.

The logarithmic behavior (\ref{logT}) is instead characteristic of a coupling constant, $\a$ for $C^2$ in this case, requiring UV renormalization,
analogously to the electromagnetic coupling in QED. Although the corresponding tensor $^{(C)\!}H^{ab}$ is traceless, and there is no trace anomaly 
in the spin-2 sector in the two-point function $\cS_2^{abcd}$, the coefficient of the logarithmic running of (\ref{logT}) and the $C^2$ coupling 
at energies $k^2 \gg m^2$ is a breaking of scale invariance which is related to the coefficient of the $C^2$ term in (\ref{anomphi}) by the
Ward Identities for the three-point function $\cS_3^{abcdef}$ of three SET correlations~\cite{TTTCFT:2019}.

\section{Summary and Discussion: The Conformal Anomaly and EFT of Gravity}
\label{Sec:EFT}

\noindent
The main results of the analysis of the gravitational vacuum polarization presented in this paper may be summarized as follows:
\begin{enumerate}[itemsep=-1mm,leftmargin=8mm, label=(\arabic*)]
\item Derivation of the Ward Identity (\ref{WI}) for the second variation of the 1PI effective action $\cS_2^{abcd}(x,y)$, from covariant 
conservation of the SET (\ref{Tqcons}), including local contact terms;
\item Verification of this WI for $F^{abcd}(k)$ defined by (\ref{Fdef}) in momentum space, by (\ref{WIk})-(\ref{WIF}) in $n$-dimensional regularization
for a scalar field of arbitrary mass $m$ and curvature coupling $\x$, the local contact terms for which are computed in Sec.~\ref{Sec:local};
\item The gravitational polarization $F^{abcd}(k)$ expressed in terms of the two transverse projection tensors (\ref{PST}), one spin-0 and traceful,
the second spin-2 and traceless, each multiplied by a Lorentz invariant form factor polarization function (\ref{Sigk}) and (\ref{Tk});
\item Identification of UV divergences by the $1/(n-4)$ pole terms (\ref{Tpole}), (\ref{Spole}) in these form factors, and their removal in the 
$\overline{MS}$ renormalization scheme, by combining with the coefficients of the terms of the local gravitational action (\ref{Sloc}) up to dimension four; 
\item Simple modification of the $\overline{MS}$ scheme by additional finite subtractions of local terms up to dimension four,
to define physical renormalized form factors (\ref{SRTR}), and dimensionless response functions (\ref{Fphys}), (\ref{FS})-(\ref{FT}) 
that vanish in the decoupling limit $m\!\to\!\infty$;
\item Identification of the corresponding physical couplings (\ref{aRbR}) of the $C^2$ and $R^2$ terms in the effective action and their renormalization group
running with physical external momentum (\ref{arun}), (\ref{brun}); 
\item Absence of $1/(n-4)$ pole in the spin-0 form factor polarization function (\ref{Spole}) for $\x\! =\!1/6$, which being completely finite and physical, 
is independent of UV regularization and automatically satisfies decoupling as $m\!\to\! \infty$ in (\ref{FSlims}) without any subtractions or modification; 
\item A compact form for the spin-0 polarization and response function (\ref{FSconf}) for $\x \!=\!1/6$, whose finite value in the 
limit of $m\!\to\! 0$ (\ref{FSlims}) is simply related to the coefficient of the $\sq R$ term in the conformal anomaly (\ref{anomphi}), 
being twice its value;
\item The imaginary part of this spin-0 response function at $\x\! =\!1/6$ defining a spectral function (\ref{sigdef}) that satisfies the 
UV finite sum rule (\ref{sumrule}) for any $m$, determined by the conformal anomaly; 
\item This spectral function becoming  a Dirac $\d$-function at $s=0$ (\ref{sigdel}) when the scalar mass $m=0$;
\item The corresponding real part (\ref{realsig}) exhibiting a massless $1/k^2$propagating Goldstone-like collective mode in the two-particle intermediate state;
\item Completely analogous behavior of two dimensonal scalars demonstrated in Appendix \ref{App:2D} and of Dirac fermions in four dimensions
in Appendix \ref{App:Ferm}.
\end{enumerate}

The conformal anomaly, like the axial anomaly, was discovered in one-loop amplitudes by regularization methods designed to manage and remove the UV divergences
of QFT, such as dimensional regularization. If the mass of the field $m\!=\!0$ and the classical action is conformally invariant, the UV and IR behvavior
of anomalous amplitudes cannot easily be distinguished. When non-zero $m$ is considered, and decoupling of the amplitude in the $m\!\to\!\infty$ limit is
preserved, so that the entire amplitude vanishes in that UV limit, the appearance of the anomaly as an IR effect in the opposite $m\!\to\!0$ limit
becomes clear. In that case $1/m$ acts as an infrared cutoff, realizing the Wilsonian IR flow to the EFT at large distances and low energies as $m$ is decreased.

Adopting this strategy and applying it to the two-point gravitational vacuum polarization function $\lag T^{ab}(x) T^{cd}(y) \rag$ exhibits 
the $\sq R$ conformal anomaly coefficient unambiguously as a genuine IR effect for scalars with curvature coupling $\x\! =\!1/6$, in 
the limit $m \!\to\! 0$, as well as for fermions for which there is no $\x$ parameter. This conclusion appears to conflict with the view
sometimes expressed in the literature that since the $\sq R$ term in the trace (\ref{anomphi}) at $m\!=\!0$ can be altered or removed entirely 
by supplying an additional local $R^2$ term to the effective action, the $\sq R$ term is scheme dependent and not part of the `true' 
anomaly~\cite{DesDufIsh:1976,Duff:1977}.

It is certainly true that one can always add a local $R^2$ term to the gravitational action (\ref{Sloc}) that would alter the $\sq R$ term in the trace.
In that case, such an addition would take the flow of the $\lag T^{ab}(x) T^{cd}(y) \rag$ effective coupling in the spin-0 sector to a finite non-zero
value at $m^2\!\to\! \infty$, which if it is added to the quantum one-loop contribution, would violate decoupling. Rather, it seems that the view
consistent with decoupling is that any finite addition to the local $R^2$ term in the gravitational action is a purely UV addition,
which explicitly violates conformal invariance and should not be confused with the one-loop amplitude of QFT, which does properly vanish
as $m^2\!\to\!\infty$. That this anomalous (rather than explicit) contribution to $\lag T^a_{\ a}\rag$ has nothing whatever to do with 
UV regularization, counterterms or renomalization is clear from the scalar $\xi\!=\!1/6$ case and that of the massive fermion treated in 
Appendix \ref{App:Ferm}, both of which are completely finite.

That these quantum loop contributions decouple as $m\!\to\!\infty$ is exactly the behavior originally postulated for UV finite amplitudes in Ref.~\cite{AppCar:1975}.
One is free to add a local $R^2$ term, viewed as a marginal deformation in the UV~\cite{AsorGorShap:2004}, but the $m^2$ dependence of the
one-loop amplitude, the UV finite spectral sum rule, and limiting form of spectral function of a Dirac $\d$-function at zero center-of-mass threshhold
unambiguously establish the $\sq R$ term in the conformal anomaly as an IR effect. Since its appearance is forced by the requirement of
the non-anomalous WI of covariant conservation, it is on the same footing as in the case of the Adler AVV axial anomaly, forced
by the WI of $U(1)$ gauge invariance in QED, and independent of regularization scheme. In that sense it is `universal,' provided only 
the WI is not violated~\cite{GiaEM:2009}.  

That the finite contributions to the $\sq R$ coefficient of the conformal anomaly are exactly $2/3$ of the coefficient of the $C_{abcd}C^{abcd}$ term
in (\ref{anomphi}) and (\ref{anomf}) also constitutes a derivation of this relation, independent of reliance on a $1/(n-4)$ counterterm in the effective action
in dimensional regularization~\cite{Duff:1977}. Rather the results of this paper suggest that this $2/3$ relation should follow from the UV finite anomalous
Ward Identities, independently of UV regularization scheme, provided only that no additional {\it ad hoc} violation of conformal invariance is introduced.
 
The main conclusion of the analysis of one-loop gravitational polarization is that the $\sq R$ term of the conformal anomaly is a genuine IR effect
that should be retained in the Wess-Zumino (WZ) effective action for the conformal anomaly in the low energy effective theory of gravity~\cite{EMEFT:2022}. 

\vspace{1cm}
\centerline{\large{Acknowledgements}}
\vspace{3mm}
It is a pleasure to acknowledge discussions with Mark W. Paris of the Theoretical Division of Los Alamos National Laboratory, and Joachim Pomper
of the Univ.~of Pisa, as well as I. L. Shapiro for bringing Refs.~\cite{BuchShap} and \cite{AsorGorShap:2004} to the author's attention, and M. Asorey and
Mark Paris for their critical reading of a preliminary draft of this manuscript.

\vspace{1cm}

\appendix
\section{Spectral Sum Rule and Massless Anomaly Pole in Two Dimensions}
\label{App:2D}

Since the integral representations of the form factors $F_i(k^2)$ are given in (\ref{Fitot}) for arbitrary $n$ dimensional flat space, one can easily
extract the $n\!=\!2$ dimensional results to compare and contrast with the $n\!=\!4$ case considered at length in the main text. 

One should recognize first that the tensor projection operators of (\ref{PST}) obeying (\ref{projs}), are $n$-dependent, according to \cite{AndMolEM:2003}
\vspace{-4mm}
\bes
\begin{align}
P^{(S)\,abcd}(k) &= \sdfrac{1}{3}\, \th^{ab}\th^{cd}= \sdfrac{1}{n-1}\, \left( \t_1^{abcd} - \sdfrac{1}{k^2} \t_3^{abcd}
+ \sdfrac{1}{k^4}\t_5^{abcd}\right)\\[2pt]
P^{(T)\,abcd}(k)&= \sdfrac{1}{2}\,\Big(\th^{ac}\th^{bd} + \th^{ad}\th^{bc}\Big) - \sdfrac{1}{n-1}\, \th^{ab}\th^{cd}\nn
&\hspace{-1cm}=\frac{1}{2} \left(\t_2^{abcd}- \sdfrac{1}{k^2} \t_4^{abcd} + \sdfrac{2}{k^4} \t_5^{abcd}\right)
- \sdfrac{1}{n-1}\, \left( \t_1^{abcd} - \sdfrac{1}{k^2} \t_3^{abcd} + \sdfrac{1}{k^4}\t_5^{abcd}\right)\label{PTn}
\end{align}
\label{PSTn}
\ees

\vspace{-1cm}
\noindent
and that for $n=2$ the tensors $\t_i^{abcd}$ are not all linearly independent. In fact
\be
\frac{1}{2} \left(\t_2^{abcd}- \sdfrac{1}{k^2} \t_4^{abcd} + \sdfrac{2}{k^4} \t_5^{abcd}\right)
= \left( \t_1^{abcd} - \sdfrac{1}{k^2} \t_3^{abcd} + \sdfrac{1}{k^4}\t_5^{abcd}\right)\qquad {\rm for}\qquad n=2\,,
\label{PT2}
\ee
so that the spin-2 projector $P^{(T)\,abcd}(k)$ in (\ref{PTn}) vanishes identically in two dimensions. This is due of course to the
fact that there are no spin-2 gravitational modes in $1+1$ dimensions, or equivalently due to the fact that the Weyl tensor $C^{abcd} =0$
vanishes identically in $n\!=\!2$. Since $P^{(T)\,abcd}$ is related to the second variation of $C_{abcd}C^{abcd}$, as follows from (\ref{varHC})
it also vanishes identically for $n\!=\!2$.

Since the Ward Identity (WI) (\ref{WI}), and relations (\ref{WIF}) hold in general $n$ dimensions, (\ref{S2F1F2}) with (\ref{PT2}) then gives
\vspace{-2mm}
\be
F^{abcd}(k)\big\vert_{n=2} =\Big( F_1(k^2) + 2 F_2(k^2)\Big) \left( \t_1^{abcd} - \sdfrac{1}{k^2} \t_3^{abcd} + \sdfrac{1}{k^4}\t_5^{abcd}\right) 
= \Big( F_1(k^2) + 2 F_2(k^2)\Big)\, \th^{ab}\th^{cd}
\label{Fn2}
\ee
so that only the spin-0 polarization survives in $n\!=\!2$ dimensions.

Secondly, since in $n\!=\!2$ the curvature coupling $\x$ can be set to zero for the scalar $\f$ field, one obtains from (\ref{Fitot})
\vspace{-3mm}
\be
F_1(k^2) + 2 F_2(k^2) =  (4\p)^{\!-\medmath{\sdfrac{n}{2}}}\, 
\G\left(2 - \sdfrac{n}{2}\right)\! \int_0^1\! dx\,  \big(M^2\big)^{\medmath{\sdfrac{n}{2}\!-\!2}}\, (2 k^4) \, x^2 (1-x)^2
\label{F1F2}
\ee
in which $n$ may immediately be set equal to $2$, since (\ref{F1F2}) is finite for this value. Thus it may be evaluated directly as
\beo
\Big[F_1(k^2) + 2 F_2(k^2)\Big]_{n=2}=  k^4 F^{(S)}\left(\sdfrac{k^2}{m^2}\right)\bigg\vert_{n=2}
\ee
with
\begin{align}
F^{(S)}\left(\sdfrac{k^2}{m^2}\right)\bigg\vert_{n=2} &=
\frac{1}{2 \p}  \int_0^1\! dx\,  \frac{x^2 (1-x)^2}{k^2 x(1-x) + m^2}
= \frac{1}{12\p k^2}\left\{1 - 6m^2 \int_0^1 \frac{x(1-x)} {k^2 x(1-x) + m^2}\right\}\nn[6pt]
& =  \frac{1}{12\p k^2}\left\{1 - \frac{6m^2}{k^2}\left[1 - \frac{2m^2}{k^2} \frac{1}{z} \ln \left(\frac{z+1}{z-1}\right)\right]\right\}
\label{F1F2k}
\end{align}
where $z$ is defined as before in (\ref{ffn}). Hence from (\ref{Fn2}) and (\ref{F1F2k}), 
\beo
F^{abcd}(k)\big\vert_{n=2} = \frac{1}{12\p k^2} \left(k^2 \h^{ab} - k^ak^b\right) \left(k^2 \h^{cd} - k^ck^d \right)
\left\{1 - \frac{6m^2}{k^2}\left[1 - \frac{2m^2}{k^2} \frac{1}{z} \ln \left(\frac{z+1}{z-1}\right)\right]\right\}
\label{Ftot2}
\eeo
for the gravitational polarization of a minimally coupled scalar field in $n\!=\!2$ dimensions.

From the first integral representation of (\ref{F1F2k}), it is clear that this finite spin-0 form factor vanishes as $m^2\!\to\!\infty$.
Indeed the function in curly brackets in (\ref{Ftot2}) has the limits
\be
1 - \frac{6m^2}{k^2}\left[1 - \frac{2m^2}{k^2} \frac{1}{z} \ln \left(\frac{z+1}{z-1}\right)\right] \to 
\left\{\begin{array}{ccc}{\displaystyle \sdfrac{1}{5^{\vphantom{(}}} \sdfrac{k^2}{m^2}}&\to 0\,,
\quad& m^2\to \infty\\[4pt]
1 , & &m^2 \to 0\end{array}\right.
\ee
so that in the first limit decoupling of the scalar loop is satisfied, while in the second limit of zero mass
\beo
F^{abcd}(k)\big\vert_{n=2} \to  \frac{1}{12\p\, k^2} \left(k^2 \h^{ab} - k^ak^b\right) \left(k^2 \h^{cd} - k^ck^d \right)
\quad {\rm for} \quad m^2 \to 0
\label{Ftot2lim}
\eeo
exhibits a massless scalar Goldstone pole. 

The residue coefficient of this pole is directly related to the conformal trace anomaly, which in $n\!=\!2$ is \cite{DavFulUnr:1976,ChrFul:1977,Duff:1977,BirDav}
\be
\big\lag \hat T^a_{\ a}\big\rag_R\, \big\vert_{n=2}= \frac{1}{24\p} \, R
\label{anomphi2}
\eeo
since by a calculation exactly parallel to (\ref{vartr})-(\ref{Siganom}), the Fourier transform
\be
\int \!d^2x\, e^{ik\cdot (x-y)} \frac{\d\, \big\lag \hat T^a_{\ a}(x)\big\rag_R}{\d g_{cd}(y)}\bigg\vert_{n=2}= 
\int \! d^2x\,  e^{ik\cdot (x-y)}  \frac{1}{24\p}\left( \frac{\d R(x)}{\d g_{cd}(y)}\right)
=  \frac{1}{24 \p} \left(k^2 \h^{cd}-k^c k^d \right)
\label{vartr2}
\eeo
around flat space, and
\beo
\frac{1}{2} \h_{ab}\,F^{abcd}_R(k)\big\vert_{n=2, m=0}= \frac{1}{24\p}\left(k^2\h^{cd} - k^c k^d \right) 
\label{F2anom}
\eeo
coincide. Thus the coefficient of (\ref{Ftot2}) in the $m\!\to\!0$ limit is precisely  twice that of the $R$ term in the conformal anomaly 
(\ref{anomphi2}) in $n\!=\!2$ dimensions, just as it is in $n\!=\!4$ dimensions, in each case given by the residue of the $1/k^2$ pole.
The main difference between the two cases is that the $1/k^2$ pole itself appears in the two-point polarization function (\ref{Ftot2})
in two dimensions, whereas only the residue appears in this two-point correlation function in $n\!=\!4$, due to the $\sq R$ form
of the anomalous term in (\ref{anomphi}), which results in an additional factor of $k^2$, compared to just $R$ in (\ref{anomphi2}).

The spin-0 polarization (\ref{F1F2k}) in $n\!=\!2$ also can be written in the dispersive form
\be
F^{(S)}\left(\sdfrac{k^2}{m^2}\right)\bigg\vert_{n=2} = \int_0^\infty\! \frac{ds}{k^2 + s}\, \s_{2D}(s)
\label{FS2D}
\ee
with
\bet
\s_{2D}(s) = \frac{1}{2 \p} \int_0^1\! dx\, x(1-x)\  \d \left(s - \sdfrac{m^2}{x(1-x)}\right)=
\frac{m^4}{\p s^3}\,\left(1 - \sdfrac{4m^2}{s}\right)^{- \frac{1}{2}}\Th (s-4m^2)
\label{sig2D}
\eeo
which satisfies the UV finite sum rule
\be
\int_0^\infty\! ds \, \s_{2D}(s)  = \frac{1}{12\p}
\label{sum2D}
\ee
determined by the conformal anomaly. As in the $n\!=\!4$ case discussed in Sec.~\ref{Sec:SpecSum}, the spectral function $\s_{2D}(s)$ becomes
a more and more sharply peaked function of $s$ near the threshhold value of $4m^2$, which falls off rapidly as $1/s^3$ for $s \to \infty$.
At $m\!=\!0$, $\s_{2D}$ vanishes pointwise for all $s >0$, and the sum rule (\ref{sum2D}) is saturated by $\s_{2D}(s)$ becoming the Dirac
$\d$-function,
\be 
\s_{2D}(s)\  \stackrel{m\to 0}{\longrightarrow}\  \frac{1}{12\pi}\, \d(s)
\ee
as follows immediately from the integral representation (\ref{sig2D}) in this limit. 

The $\d$-function in the spectral function in the massless limit signals the appearance of the massless collective excitation and $1/k^2$ pole 
in the real part of the response function (\ref{FS2D}), by
\be
F^{(S)}\left(\sdfrac{k^2}{m^2}\right)\bigg\vert_{n=2}\  \stackrel{m\to 0}{\longrightarrow}\  \frac{1}{12\p} \frac{1}{k^2}
\ee
which is also manifest in the 1PI effective action 
\be
\cS[g] = -\frac{1}{96\p} \int\!d^2\!x\!\sqrt{-g(x)} \!\int\! d^2\!y\!\sqrt{-g(y)} \ R(x)\, \big(\!\sq^{-1}\big)_{xy} \, R(y)
\label{Poly2D}
\ee
of the conformal anomaly in two dimensions~\cite{Poly:1981,Poly:1987}. This massless collective excitation may also be represented
as the propagator of an explicit scalar degree of freedom $\vf$ (conformalon), in the local effective action
\beo
S_{\!\cA}^{\rm 2D}[g;\vf]= - \frac{1}{96\p}\int\!d^2\!x\!\sqrt{-g}\, \left\{g^{\m\n} (\na_\m \vf )(\na_\n \vf)- 2R \vf\right\}\,.
\label{2Danom}
\eeo
of the 2D conformal anomaly (\ref{anomphi2})~\cite{BlaCabEM:2014,EMEFT:2022}.

The dispersion relation, and UV finite spectral sum rule for the gravitational vacuum polarization $\lag TT\rag$ of Dirac fermions in two dimensions has been
considered by the authors of \cite{BertlKohl:2001}, with similar results to the scalar case.

\section{Spectral Function and UV Finite Sum Rule for Fermions}
\label{App:Ferm}

The gravitational vacuum polarization of Dirac fermions in four dimensions has been calculated previously in Ref.~\cite{GorShapII:2003}. 
The finite spin-0 response function for fermions can be expressed 
\be
F^{(S)}_f\left(\sdfrac{k^2}{m^2}\right) = k^2\! \int_0^\infty\! \frac{ds}{k^2 +s}\, \s_f(s)
\label{FSfsig}
\eeo
analogous to that of Secs.~\ref{Sec:Phys}- \ref{Sec:SpecSum} for scalars, with 
\beo
\s\!_f(s) = \frac{1}{48\p^2}\int_0^1 \! dx \, (1-2x)^4 \ \d \left(s - \sdfrac{m^2}{x(1-x)}\right)=
\frac{m^2}{24\p^2 s^2}\,\left(1 - \sdfrac{4m^2}{s}\right)^{ \frac{3}{2}}\Th (s-4m^2)\,.
\label{sigf}
\eeo
This gives several useful integral representations for this function, and the final result
\begin{align}
&F^{(S)}_f\left(\sdfrac{k^2}{m^2}\right) = \frac{k^2}{48\p^2} \!\int_0^1 \! dx\,\frac{x(1-x) (1-2x)^4}{k^2 x(1-x) + m^2}
= \frac{1}{16\p^2}\,\frac{1}{15} \left\{ 1 - m^2 \!\int_0^1\! dx\,\frac{(1-2x)^4}{k^2x(1-x) + m^2}\right\}\nn[6pt]
&= \frac{1}{48\p^2} \!\int_0^1 \! dx\, x(1-x)(1-2x)^3 \,  \frac{\pa}{\pa x} \ln \left(\sdfrac{M^2}{m^2}\right)
= \frac{1}{48\p^2} \!\int_0^1\, dx \, \ln \left(\sdfrac{M^2}{m^2}\right) (1-2x)^2\, (10x - 10x^2 -1)\nn[6pt]
&\hspace{2cm}= \frac{1}{16\p^2}\,\frac{1}{15} \left\{1 + \frac{20}{3} \frac{m^2}{k^2} 
- 10\,\frac{m^2}{k^2} \, z^2 \left[z \, \ln\left(\sdfrac{z+1}{z-1}\right) -2 \right] \right\}
\end{align}
with $z$ is defined by (\ref{ffn}). This function has the limits
\be
F^{(S)}_f\left(\sdfrac{k^2}{m^2}\right)  \to \frac{1}{16\p^2} \frac{1}{15} 
\left\{\begin{array}{cclc}{\displaystyle \sdfrac{1}{14^{\vphantom{(}}} \sdfrac{k^2}{m^2}}\to 0&,&\ m^2\to \infty\\[4pt]
1 &,&\ m^2 \to 0\end{array}\right.
\label{FSflims}
\ee
satifying decoupling in the large $m$ limit. On the other hand, the conformal anomaly for massless fermions is~\cite{DowkCritch:1977,Duff:1977,BirDav}
\be
\big\lag \hat T^a_{\ a}\big\rag_R\,\Big\vert_{f,m=0}= \frac{1}{16\p^2}\left\{- \frac{11}{360}\,E + \frac{1}{20} C_{abcd}C^{abcd} +  \frac{1}{30}\sq R\right\}
\label{anomf}
\ee
so that in the opposite $m\!\to\!0$ limit $F^{(S)}_f$ gives exactly twice the coefficient of the $\sq R$ term in the anomaly, as in (\ref{vartr})-(\ref{Siganom}) 
of the text for conformal scalars.

The fermion spectral function (\ref{sigf}) also satisfies the UV finite sum rule
\be
\int_0^\infty\! ds \, \s\!_f(s)  = \frac{1}{16\p^2}\, \frac{1}{15} = \frac{1}{240\p^2}
\label{sumf}
\ee
for any $m$, and becomes the Dirac $\d$-function distribution
\be
\s\!_f(s)\  \stackrel{m\to 0}{\longrightarrow}\  \frac{1}{16\p^2}\, \frac{1}{15}\, \d(s)
\label{sigfdel}
\ee
as follows immediately from the integral representation (\ref{sigf}) in this limit. The coefficient in (\ref{sumf}) and (\ref{sigfdel}) is again
that of the $m\!\to\!0$ limit of (\ref{FSflims}), determined by the conformal anomaly for fermions. 

Thus, the results and conclusions presented in the text regarding the infrared nature of the $\sq R$ term of the conformal anomaly for scalars for 
$\x\!=\!1/6$ apply to the completely UV finite and regularization independent spin-0 gravitational polarization of Dirac fermions as well. 

\pagebreak
\centerline{\large{\bf References}}
\bibliographystyle{apsrev4-1}
\vspace{-5mm}
\bibliography{../gravity}

\end{document}